Graphical Abstract

# Nanowire: Anisotropic initial reshaping and breakup dynamics

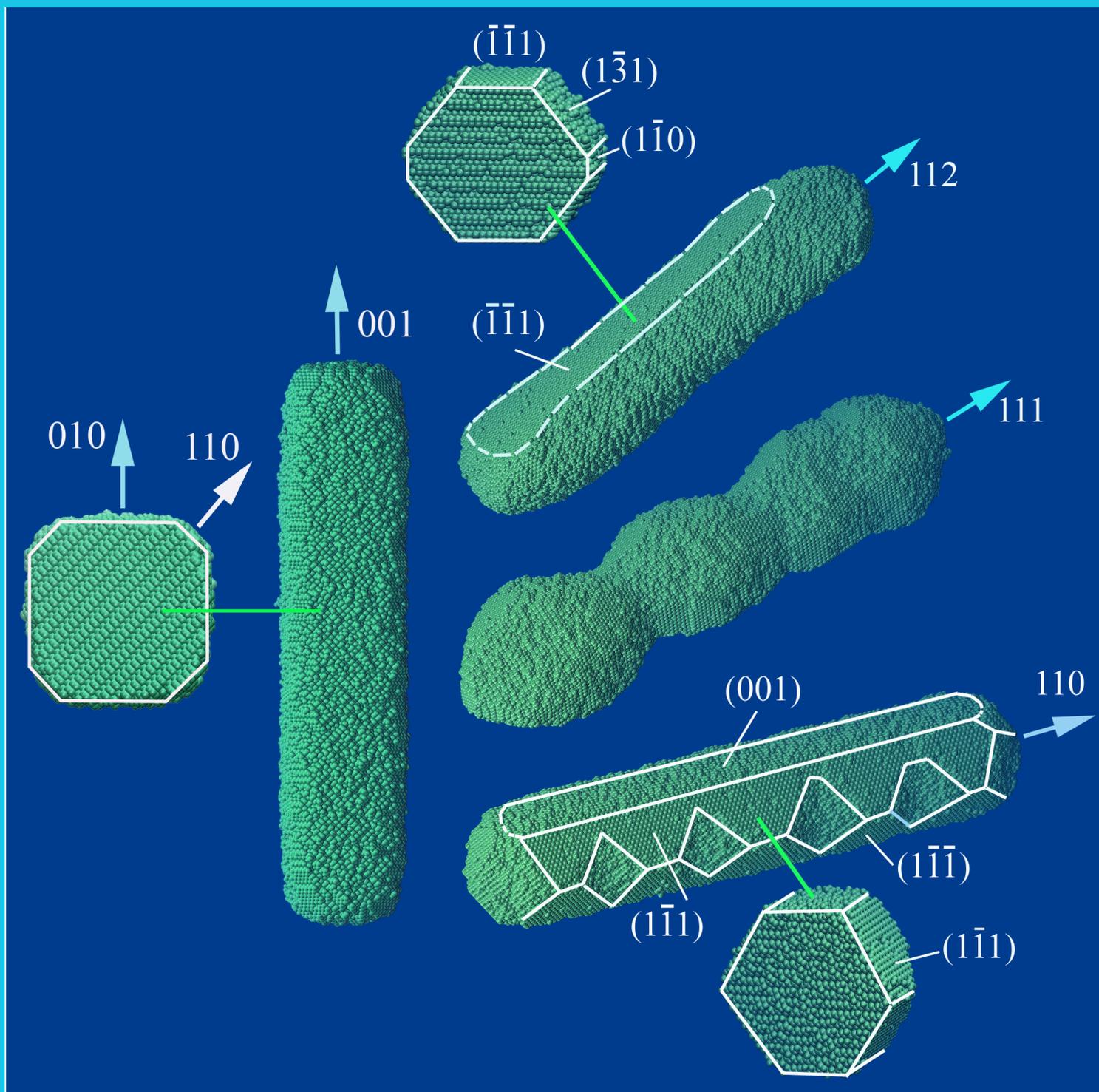
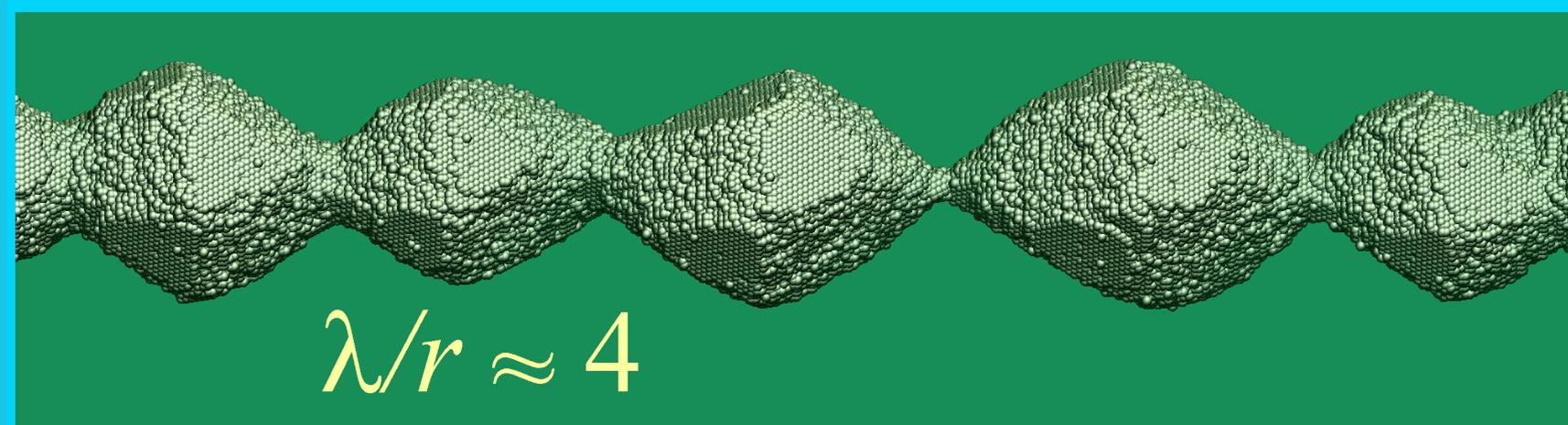
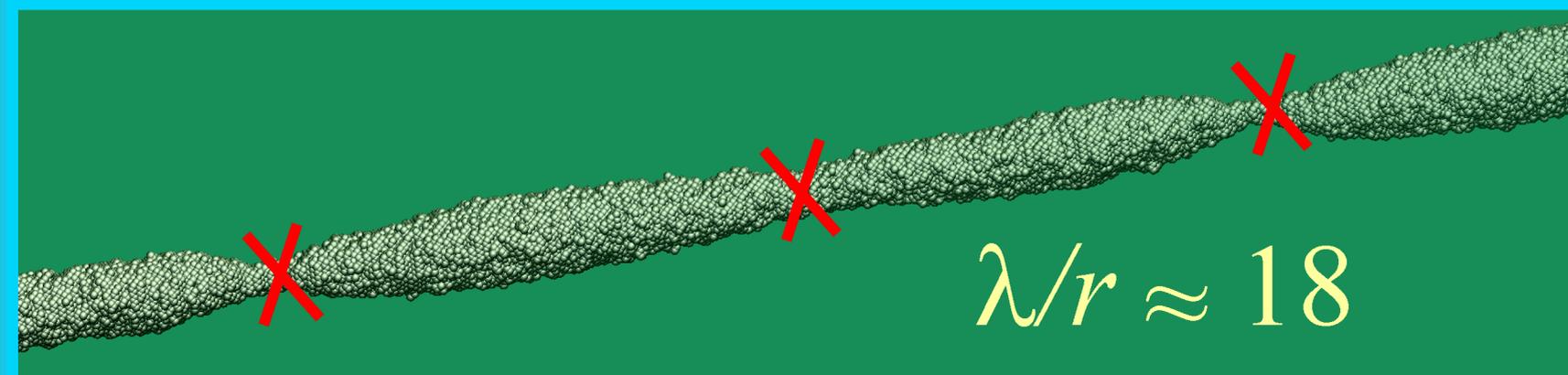
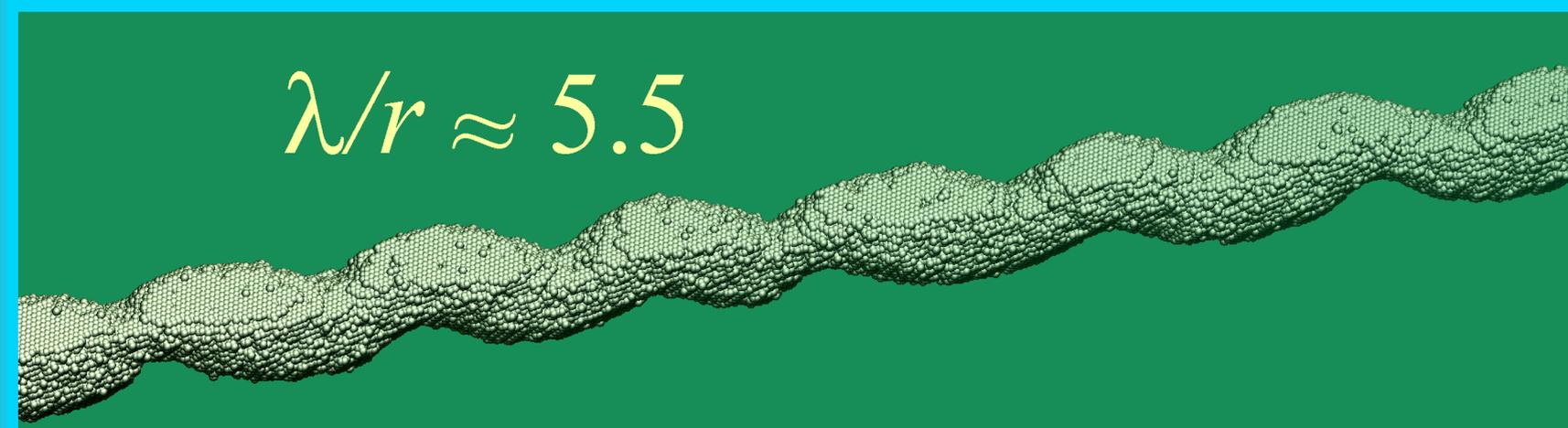

# Restructuring and Breakup of Nanowires with the Diamond Cubic Crystal Structure into Nanoparticles

Vyacheslav N. Gorshkov[1], Vladimir V. Tereshchuk[1], and Pooya Sareh[2*]

[1]National Technical University of Ukraine, Igor Sikorsky Kyiv Polytechnic Institute, 37 Prospect Peremogy, Kiev 03056, Ukraine.

[2]School of Engineering, University of Liverpool, London Campus, 33 Finsbury Square, London EC2A 1AG, United Kingdom.

*Corresponding author. Email: pooya.sareh@liverpool.ac.uk

**Abstract**. A kinetic Monte Carlo approach is applied to study physical mechanisms responsible for the breakup of nanowires with the diamond cubic crystal structure into a chain of nanoparticles discovered in preceding experiments on Silicon nanowires. We show that this process is based on the well-known mechanism of roughening transition, which specifically manifests itself in quasi-one-dimensional systems/nanowires with a pronounced anisotropy of the surface energy density. Depending on the temperature and orientation of the nanowire relative to its internal crystal structure, the wavelengths of substantial cross-sectional modulations exceed its initial radius by 4 to 18 times. For certain orientations, a straight nanowire at the initial stage of evolution forms a serpentine/helical structure. The scenarios of the stage of nanowire ruptures into single nanoclusters are also diverse: either each spindle-shaped region of the nanowire transforms into a separate drop (by long-wave surface perturbations), or the adjacent short-scale beads absorb each other due to the Ostwald ripening effect, which can be accompanied by the formation of long-lived many-body dumbbells. The discovered features of the dynamics of quasi-one-dimensional systems expand our conceptions of the physical mechanisms involved in the breakup of nanowires (presented by Nichols and Mullins as a classical model for such instabilities) which could be useful in applications based on chains of ordered nanoparticles.

# 1. Introduction

Si- and Ge-based semiconductor nanowires have found numerous applications in recent years as a result of their outstanding physical properties. These include large intrinsic carrier mobility[1-3], suppressed phonon thermal conductivity[4], and low resistivity[5,6], making them promising building blocks for optical and nanoelectronic devices[7-11]. For example, modern methods of synthesis of core-shell semiconductor nanowires allow fabrication of diameter modulated nanochain heterostructures[12-15] which are nowadays used in electrochemical energy storage and solar energy conversion[16,17]. Moreover, it was reported[18-20] that silicon nanowires play a key role in the construction of sensitive biosensors capable of diagnosing cardiovascular diseases and various types of cancer.

Despite such favorable characteristics, the thermal stability of Si- and Ge-based semiconductor nanowires and their morphological evolution under elevated temperatures have recently been of concern[21-29]. In general, since nanowires have small size and large area-to-volume ratio, their physical properties differ from those of bulk materials, they tend to form bulges and eventually break-up into isomeric nanoparticles at the premelting temperatures[25-29,30-35]. Although this effect can be favorably exploited in the design of optical waveguides[36-37], naturally, it also impairs the optoelectrical properties of the system. Therefore, it is of crucial importance to study the dynamics of the break-up process since it may provide valuable insights into how to tune the morphological features of nanowires.

Diameter-modulated nanowires can be produced in a number of different ways. This presented study concerns two of them which are based on clear physical mechanisms and may be theoretically analyzed in simple limiting cases.

The first one corresponds to the analogue of the Plateau–Rayleigh instability[38] observed in cylindrical liquid jets. The classical Nichols and Mullins model[39] considers the development of initial sinusoidal perturbations of the nanowire radius, $\Delta r(z,t) \sim \varepsilon(t) \times \sin(2\pi z/\lambda)$, due to the surface diffusion of bonded atoms from the zones with greater surface curvature to the zones with smaller surface curvature ($\varepsilon(t)$ is the modulation amplitude, $\lambda$ is the wavelength of perturbations, and $z$ is the longitudinal coordinate). The arising spontaneous modification of the nanowire shape is accompanied by a decrease in its surface (potential) energy, $E_{surf}$, which limits the spectrum of possible spatial harmonics from below: $\lambda > 2\pi r_0$ ($r_0$ is the initial radius of the nanowire).

The largest increment in time has the wavelength of perturbations of the order of $\lambda_R \approx 9r_0$, which corresponds to the optimal ratio between the displaced mass and the accompanied change in surface energy $\Delta E_{surf}$. In this model, it is assumed that the surface energy density, $\sigma$, is isotropic, and the shape of the nanowire corresponds to the increase of sinusoidal modulation, $\varepsilon(t)$, of its radius in time. These assumptions limit the application of this model to the interpretation of experimental results for a number of reasons pointed out below.

The second option for creating a diameter-modulated nanowire is based on the unstable growth of the shell during the diffusion deposition of material on the nanowire[14] with various modifications of the accompanying physicochemical features of the process. In this case, the surface transformation may be qualitatively represented on the basis of Mullins and Sekerka model[40] of surface instability, and its description is based on transparent physical mechanisms. The formation of a periodically modulated shell on the nanowire, $\Delta h(z,t) \sim \delta(t) \times \sin(2\pi z/\lambda)$, causes a redistribution of diffusion fluxes on its surface. The ratio of the diffusion flux densities on the nanowire surface in the thickening regions, $\Phi^{(+)}$, and the narrowing regions, $\Phi^{(-)}$, depends on the wavelength of sinusoidal perturbations, $\lambda$. The condition for increasing their amplitude, $\delta(t)$, ($\Phi^{(+)} > \Phi^{(-)}$), determines the range of relatively short wavelengths, $\lambda < 2\pi r_0$, (see section S1 in the Supplementary Information) in contrast to the Plateau–Rayleigh instability[38]. The rate of surface diffusion of bonded atoms, which suppresses short-scale radius modulations, assigns the minimum value, $\lambda_{min}$, at a given level of supersaturation of the material being deposited.

In this paper, we study the break-up patterns of nanowires with the diamond cubic crystal structure (i.e., Ge and Si). In contrast to the conventional understanding of the break-up process, anomalously short-wave sausage-like nanowire configurations and abnormally narrow gaps, $\Delta$, between adjacent nanodroplets were observed in a series of experiments[12,13]. As can be seen in Fig.1a, the ratio $\lambda/r_0 \sim 4.2$ is much smaller than the critical value $2\pi$ at the stage of substantial modulation of the nanowire radius. Moreover, Fig. 1c implies that $\Delta/d < 1$ (rather than $\Delta/d \approx 1.38$ according to the classical model[39]), where $d$ is the average initial diameter of nanoclusters.

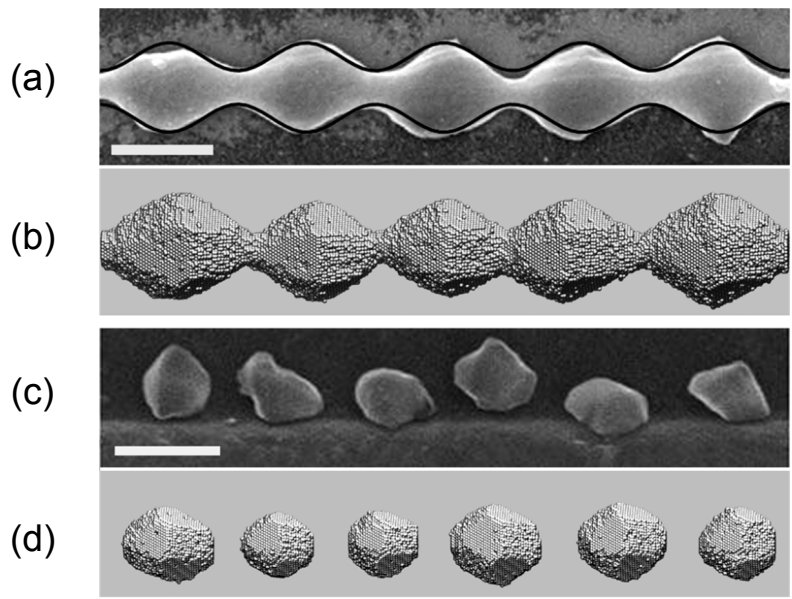

**Fig. 1.** (a) and (c): Experimental results [12] that depict the self-transformation of a straight Silicon nanowire into a continuous or discrete island-chain nanowire structure (white scale bars represent 200 nm). The black curve in part (a) is our approximation for the experimentally observed dependency of the nanowire radius on its length. (b) and (d): Results of our Monte-Carlo model which show (b) the formation of the long-lived quasi-stable state of a cylindrical Si-nanowire with an initial diameter $d_0 \approx 7.7\ nm$, and (d) the break-up of a nanowire with $d_0 \approx 6.2\ nm$ into a nanodroplet chain.

Some of the results of our numerical experiments are also depicted in Fig. 1b and d for comparison with the experimental data[12]. The model we use does not have the disadvantages of a linear approximation, unlike the mentioned classical model[39], where (i) the surface energy density is isotropic, (ii) the nanowire surface modifications are sinusoidal, and (iii) the exchange of the nanowire surface with the free atoms surrounding it, by means of the detachment-reattachment processes, is not taken into account.

The last of the aforementioned factors strongly determines the discussed features of the nanowire breakup. Under such conditions, we associate the short-wavelength perturbations of the surface of considerable amplitude (Fig. 1) with the well-known effect of the roughening transition[41-47]. In fact, this phenomenon resembles the case of unstable diffusion growth (see above) which is self-consistent with its own vapor. Note that the surface energy of a substantially modified nanowire (Fig. 1b) exceeds the surface energy of a wire of the same volume with a constant cross-section at the fixed length. This does not contradict the laws of thermodynamics, since the increase in the surface energy of the nanowire, $\Delta E_{surf}$, is accompanied by the increase in entropy, $\Delta S$, so that the change in free energy would be $\Delta F = \Delta E_{surf} - T\Delta S < 0$ at a given temperature, $T$. Naturally, $\Delta E_{surf}$ begins to decrease with further evolution of the nanowire with the formation of ruptures. It is this stage, when $\partial \Delta E_{surf}/\partial t < 0$, that should be referred to as the Plateau–Rayleigh instability[38]. We have shown that the diversity of break-up scenarios strongly depends on the temperature of the nanowire and on the orientation of its axis relative to the internal crystal structure. For instance, in contradiction to the results shown in Fig. 1 (the nanowire axis is oriented along the [111]-direction), the abnormally long-wave perturbations ($\lambda/r_0 \sim 18$) result in breaking the nanowire when its axis is oriented along the [100]-direction.

We have also found out the possibility of the formation of helicoidal fragments, extended snake-like configurations, and many-body long-lived structures (see 'Movie S1' and 'Movie S2' in the Supplementary Information). In conclusion, we note that the disintegration processes of nanowires with the diamond cubic crystal structure are strikingly different from the break-up processes of nanowires with the FCC (face centered cubic) lattice structure[30-33], even though the Wulff configurations[48,49] for these types of crystal lattices are visually very similar.

## 2. Material and methods

The present MC approach was developed[50,51] to model shape selection in fast synthesis by growth from highly supersaturated solutions of isomeric nanocrystals for catalysis, sintering, and other applications[57]. The model was later used to model the surface- and nanotube-templated growth[53,56] of nanoparticles and other nanostructures, the sintering of the former,[54,55] as well as the breakup of nanowires.[32,33] Recent studies have primarily considered the FCC symmetry of noble-metal materials as they are of interest in the applications of the relevant nanoparticles[36]. However, here we study the diamond lattice structure of the carbon-group materials, partly because nanowire experiments[4,22,23,26] use such materials. We note that the diamond lattice can be conveniently considered as embedded in the 'diamond cubic' crystal structure units, as shown in Fig. 2a.

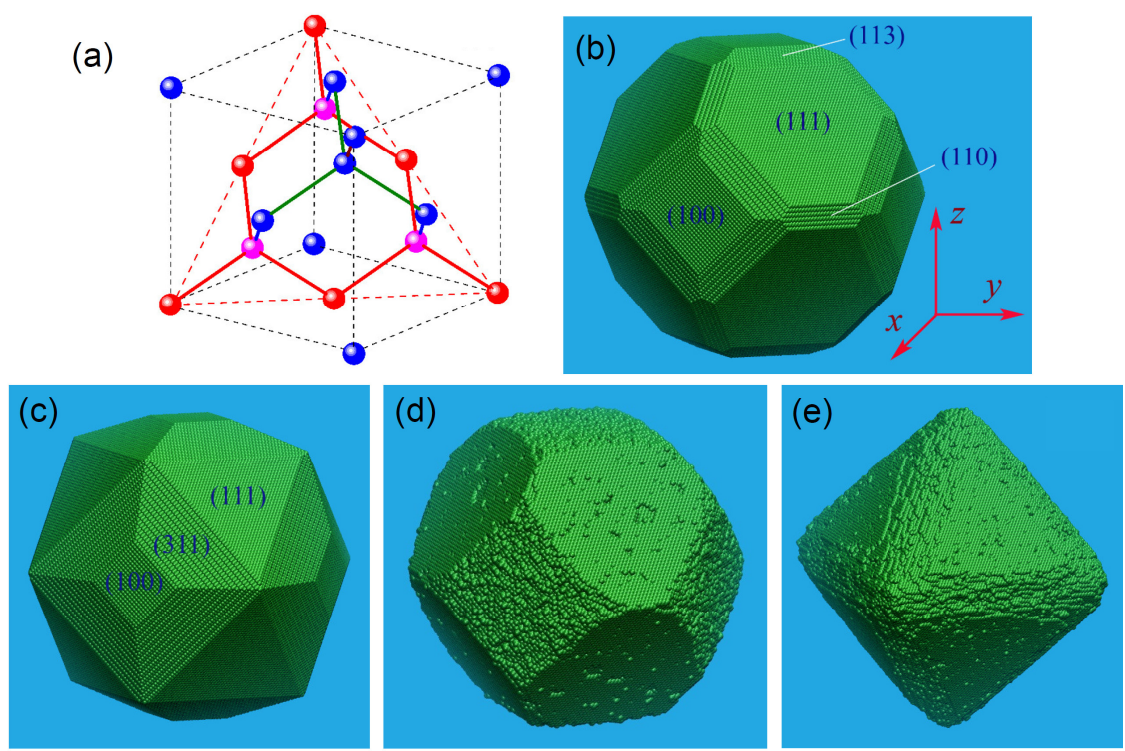

**Fig. 2. (a)** Diamond cubic representation of an 8-atom sub-unit of the diamond lattice. Solid lines represent actual bonds, whereas broken lines are illustrated for guiding the eye only. For later use, we note that the (111)-type lattice plane includes the six red-colored atoms, as well as the three magenta-colored atoms, which are here somewhat 'above' the geometrical plane of the red broken-line triangle. **(b)** & **(c)** Schematics of equilibrium-shape Wulff construction for the Si crystal based on experimental values for surface energies and for theoretically calculated values, respectively[64]. **(d)** Depiction of a steady-state crystal shape obtained using our model $\alpha = 2.7$ and $p = 0.36$. **(e)** A steady state of a nanoparticle when evaporation from its surface is intentionally blocked.

In the present model, it is assumed that nanostructure dynamics involves fast, nonequilibrium processes driven by matter imbalance. Atoms (or molecules) constitute a 'gas' of diffusers that can attach to the nanostructure. Atoms that are part of the nanostructure can hop to nearby vacant crystal sites, or detach (i.e., rejoin the 'gas'). To make the modeling numerically manageable for large structures, and for comparison with experimentally observed trends, an important assumption is made that atoms can only be attached to (or 'registered' with) the underlying lattice structure. For definiteness, consider a monoatomic structure with coordination number $m_c$ (here $m_c = 4$) and lattice spacing $\sqrt{3}a/4$, where $a$ is the cubic cell linear size (Fig. 2a). In the description of our results in the following sections, we found it convenient to measure distances in terms of

$$\ell = a/\sqrt{2} \qquad (1)$$

It is important to note that this is unrelated to the fact that this length also happens to be the next-nearest neighbor distance of the lattice, i.e. the minimal distance between the red atoms in Fig. 2a.

Atoms are represented as the diffuse 'gas' generated by off-lattice hopping at random-angles, with fixed-length steps that are a fraction of $a$ (here we take this length to be $\sqrt{3}a/4$ - unrelated to the fact that this happens to be the nearest neighbor distance of the lattice). If a gas atom hops into a Wigner-Seitz cell centered at such a vacant site, but with its nearest neighbor(s) occupied, it attaches at that lattice position. This 'precise registration' of the atoms with the crystal lattice[50-54] limits the model to morphologies of relevance to an important class of crystalline nanostructures synthesized by fast non-equilibrium processes, which includes isomeric nanocrystals.[57] Crystal atoms' dynamics will be described shortly. Such nanostructures do not have any structure-spanning defects that control their shape selection by the preferential growth of certain crystalline faces. The present model has been successful in describing shapes in the synthesis of 'isomeric' nanocrystalline particles, which are even-proportioned. In reality, large defects are dynamically avoided (or are not nucleated) at the microscopic scales. The precise registration rule phenomenologically imposes the same property in mesoscopic modeling. For a certain range of sizes, nanocrystals then grow with certain model-parameter-dependent proportions. Thermal fluctuations are present at surfaces, but overall particle shapes are bounded[50,51] by the lattice planes of symmetries similar to those in the equilibrium Wulff constructions.[58-60] Particle shapes for the SC, BCC, and FCC crystal symmetries[50,51] modeled in this way were consistent with the experiments for metal/oxide nanocrystals[61] and core-shell noble-metal nanoparticles.[62,63]

It was found that this model has also been useful in exploring the shapes and dynamics of nanostructures other than isomeric nanoparticles. These include surface-templated nanopillars[52,53] and the dynamics of bridging and merger of nanoparticles during sintering[54,55] etc. Furthermore, by enclosing the nanostructure and the 'gas' in a 'container' with a fixed concentration of atoms maintained in a thin layer as its walls, or with reflective walls, one can effectively model the surrounding medium. In the latter situation, the nanostructure can grow

(effective supply of atoms), or dissolve (effective removal of atoms), or reach a steady state with its surrounding. Here we focus on the latter regime. In spite of these limitations, the present model provides[32,33] useful information on the extent to which observed nanowire instabilities can emerge in situations with no manifest liquid-like effects (no premelting of the surface), and where the breakup is not predetermined by the dynamics of large internal defects. The model focuses on the dynamics driven by the net surface transport of matter (surface diffusion) and/or the steady-state exchange of matter (atoms) with the surrounding 'gas'.

Each crystal-lattice atom, if not fully surrounded, can hop to the vacant nearest neighbor site (the next-nearest-neighbor hopping has also been considered[50]). The probabilities of specific moves are proportional to the temperature-dependent Boltzmann factors. A unit MC time step corresponds to a random sweep through the system, such that the diffuser gas atoms are moved once on average, whereas lattice atoms have one hopping attempt on average. Movable lattice atoms have coordination numbers $m_0 = 1, \ldots, m_c - 1$ (recall that here $m_c = 4$) and they actually attempt to hop with probability proportional to $p^{m_0}$. Here we introduce the activation free-energy barrier $m_0 \Delta > 0$, and $p = e^{-\Delta/kT} < 1$. If the 'attempt to move' is carried out, the atom ends up in one of its $m_c - m_0$ vacant nearest neighbor sites, or is kept in its original site. The final positioning is selected with the probability proportional to the inverse of a free-energy change Boltzmann factor, $e^{m_t|\varepsilon|/kT}$ (normalized over all the $m_c - m_0 + 1$ targets including the original position), with $\varepsilon < 0$ measuring the free-energy of the binding at the target sites. Note that the target-site coordination in the resulting configuration can be $m_t = 1, \ldots, m_c - 1$ for hopping, and $m_t = 0$ for detachment, where in the latter case, just-detached atoms join the diffuser gas when they are next 'probed' if randomly chosen during MC sweeps.

In summary, with each unit-step MC sweep, a sufficient number of random selections is made so that on average each atom is 'visited' once. If the atom is in the gas, a random-direction diffusion hop is carried out; if the final position is within a unit cell near a crystal, this atom attaches to the lattice. The atoms already in the crystal, if not blocked, are moved with probability determined by an activation free-energy barrier; the end position is selected according to the Boltzmann factors as described earlier. The end position can be either a part of a crystal, or the move can leave the atom, and/or (some of) its neighbors, not connected to the crystal. Such disconnected atom(s) are all reclassified as located in the gas. The mobility of the surface atoms (the surface diffusion coefficient) is determined by the parameter $p$, controlled by the activation free-energy scale $\Delta$, introduced above

$$p = e^{-\Delta/kT}. \tag{2}$$

The second free-energy scale, $\varepsilon$, introduced above, reflects the local binding of crystal atoms, and we define

$$\alpha = |\varepsilon|/kT. \tag{3}$$

Previous studies[50-56] for other lattice structures with larger coordination numbers have suggested that experimentally relevant nonequilibrium morphologies are described by this model for a range of mesoscopic sizes provided we use reference values $\alpha_0$ and $p_0$ comparable to 1. Temperature, $T$, can then be increased or decreased by varying $\alpha$, which is inversely proportional to it, provided $p$ is adjusted according to

$$p = (p_0)^{\alpha/\alpha_0}. \tag{4}$$

For the present lattice structure with smaller coordination, comparing typical binding energies of various lattice planes, for instance for Si and Ge versus those for noble-metal Au (see Table 2.1) in[64], and also based on our preliminary simulations for single nanoclusters, we chose the following ranges of values

$$\alpha = 2.4 - 3.0 \quad \text{and} \quad p = 0.32 - 0.40, \tag{5}$$

and

$$\alpha_0 = 2.7 \quad \text{and} \quad p_0 = 0.36. \tag{6}$$

In actual simulations, the nanostructure is enclosed in a container. The container surface is placed at distances from the initial nanostructure that are typically 10 times larger than the structure dimensions. In all simulations here, the container was reflective, and the total number of atoms remained unchanged. After a short transient time, enough atoms detach from the initial shape to form a 'gas' in steady state with the nanostructure. Since this was a small fraction of the total number of atoms, for convenience the container was assumed initially vacant. For the diamond lattice structure considered here, we performed some preliminary simulations, which are illustrated in Figs. 2d and 2e, where we show a nanoparticle that started from a spherical shape of initial radius $45\ell$ (i.e. around 32 lattice constants $a$) cut out from the crystal structure, and then reached a steady state with the gas in a container. This simulation was performed for our reference model parameters given in Equation (6). The initial sphere (and then the entire system) contained $1.08 \times 10^6$ atoms, of which approximately $1.05 \times 10^6$ atoms remain in the particle when the equilibrium state is achieved (Fig. 2d). The shape of nanoparticles shown in Fig. 2d is similar to that shown in Fig. 2b.

Note that the kinetic processes on the surface of the nanoparticle and the exchange of its surface with the gas of free atoms can induce significant modification of its equilibrium form. It is known that natural diamonds mainly have the octahedral shape. According to the present model, this phenomenon can be realized when the sublimation of atoms from the particle surfaces is excluded. The shape depicted in Fig. 2e corresponds to a numerical simulation when evaporation from the surface is artificially blocked (if a bonded atom has obtained the 'permit' to become a free atom at some moment in time, then its detachment from the nanocluster surface is cancelled; i.e., only drift along this surface is allowed).

# 3. Results

## 3.1. Anisotropy of the nanowire initial reshaping

The results presented here demonstrate that the variety of breakup scenarios in nanowires with a diamond-like crystal structure is much wider than in the case of FCC lattice. The substantial influence of the nanowire orientation - relative to its crystal lattice structure - on the nanowire surface dynamics is shown in Fig. 3. An initially cylindrical nanorod with a length, $L$, ten times its radius, $R$, transforms to essentially different configurations at the initial stage of evolution.

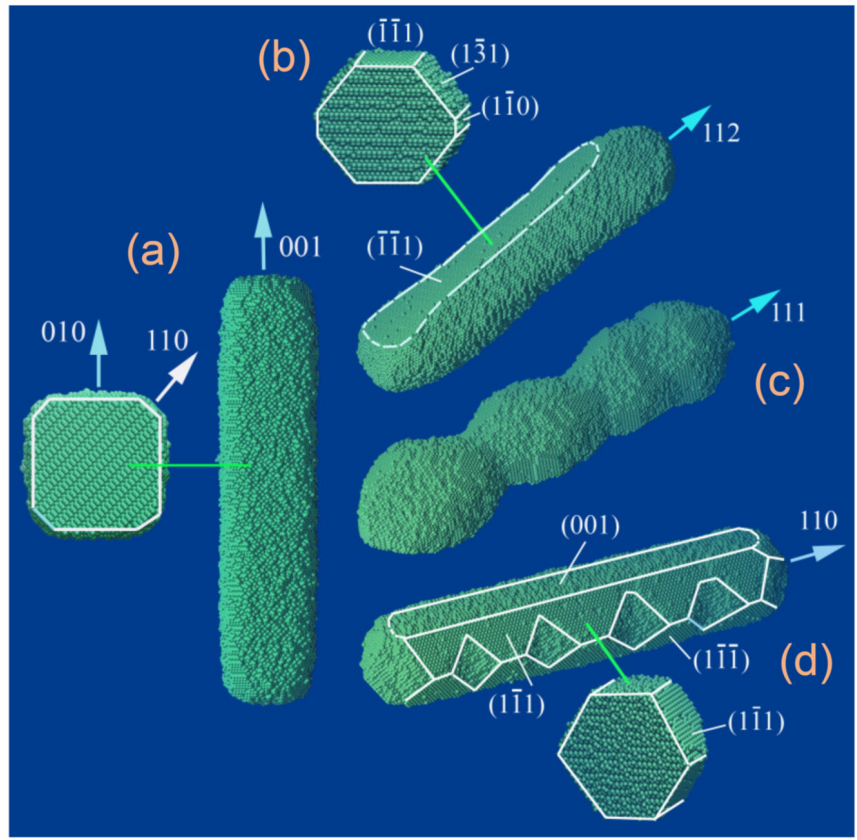

**Fig. 3.** Snapshots of configurations for four short, $L = 120$, initially cylindrical nanowires at intermediate temperature: $\alpha = 2.7, p = 0.36$, with relatively large diameter $d_0 = 24$ (~9.2 nm), evolved for $t = 7, 6, 3, 6$ ($\times 10^6$) MC time steps, for initial orientations (**a**) [001], (**b**) [112], (**c**) [111], and (**d**) [110]. For orientations [001], [112] and [110], cross-sections are cut out at the locations marked by green lines. The white outlines are added to guide the eye.

The shapes of the nanorod cross-sections in Fig. 3a, b, and d are physically expected and qualitatively understandable. The minimization of the Helmholtz free energy, $F$, at a given temperature, $T$, is accompanied by the formation of facets on the lateral surface of the rod, which are characteristics for the Wolff configuration of a single nanocluster and are parallel to its axis.

Moreover, the facets of these types ('cutting' facets) dominate in the formation of slopes at neck regions/'sausage' with the development of surface instabilities and subsequent ruptures of a long nanowire. Here the neck regions are called 'sausages', analogous to the so-called 'sausage' instability in plasma physics. Types of the arising cutting facets (of different areas and with different surface energy densities) have to provide the further decrease in the free energy, $F(t)$. These features of neck slopes formation have been discussed[32,33] for the FCC lattice structure. For instance, in the case of [001]-nanowire with diamond-like crystal structure, the neck slopes must be formed by eight facets of the [111]-type (four on the left and four on the right of the region of a rupture), as it happens in the case of FCC lattice[32,33].

However, nanowires with the diamond cubic crystal lattice exhibit a striking variety of breakup scenarios due to the specific anisotropy of their surface energy density, which forms different distributions of this density on the nanowire surface for different nanowire orientations. Note, for instance, that in the cases of [001]- and [112]- orientations, the cross-sections of these nanorods are almost homogeneous in the longitudinal direction, opposite to the nanorods with [111]/[110]-orientations (Fig. 3). Thus, as it will be shown shortly, the [001]/[112]-nanowires are characterized by an abnormally long period of surface perturbations, $\Lambda$, where we have $\Lambda/r_0 \sim 15 - 18$; this period is much longer than the well-known value $9r_0$ (see Movie S3 in the Supplementary Information). Furthermore, the breakup dynamics of the [112]-nanowire is rather unusual (see Fig. 6A and B, Fig. S6 and Fig. S11 in the Supplementary Information).

The orientation [110] demonstrates short-scale perturbations of the lateral edges (Fig. 3d) – notches created by the facets of [100]-types. The appearance of such surface modulations – the effects of surface roughening[41-46,65,69] – absolutely does not correspond to the classical model[39] for the instability under consideration. These notches may result in the nanowire breakup at high temperatures ($\alpha \sim 2.4$), which is presented and discussed in the Supplementary Information (Figs. S9 and S10). Now, we focus on the effects of roughening transition when a nanowire is oriented along the [111]-axis (Fig. 3c), and this phenomenon manifests itself most effectively.

### 3.2. Effects of roughening transition on the nanowire dynamics

Referring back to Fig. 3, the greatest modulations in the cross-section are demonstrated by the nanorod with [111]-orientation for the time that is twice shorter than in the cases of [001], [112], and [110]-orientations. The characteristic size of the created beads is comparable to the initial diameter of the nanorod. The excitation of such short-wave perturbations of the surface may be due to the so-called end-effect. The mechanism of this effect is similar to the disintegration of limited liquid filaments[68]. The excessive Laplace pressure at the ends of such filaments leads to local shrinkages near these ends and the formation of broadenings (beads) adjacent to the initially unperturbed part of the filaments. Such a transformation of the surface shape leads to the formation of a reduced Laplace pressure (of negative radii of curvature) in the zones of junction. The emerging liquid flows from the inside of the liquid filament to its ends lead to the occurrences of neck regions. In fact, capillary waves with a length of the order of the filament diameter propagate from the ends of the filament towards the opposite ends. Nonlinear effects result in a series of drops cutting off the ends of the shrinking liquid thread.

The discussed effect (end-effect) is analyzed in detail for finite nanowires with FCC crystal structures[32,33]. It is shown that the surface diffusion of atoms can also lead to perturbation waves of the nanowire surface, propagating from its ends into the inner part. The interference of these waves often gives rise to the formation of long-lived quasi-equilibrium dumbbells (analogues of Delaunay unduloids[66]). Further dynamics of the nanorod, as presented in Fig. 3c, demonstrates the possibility of being of such a three-body configuration (see Movie S2 and Fig. S8).

The end-effect is very sensitive to the orientation of the nanowire, which is associated with the anisotropy of the density of surface energy. This statement is clearly demonstrated by the simulation outputs presented in Fig. 3. The factors that inhibit the manifestation of this effect are physically simple: (i) the intense surface diffusion suppresses the formation of localized beads at the ends of the nanowire, and (ii) the formation of new surface fragments in the thickening regions increases the free energy of the nanowire. Importantly, the end-effect can break a finite nanowire into droplets in cases where the manifestation of the classical Plateau-Rayleigh instability is either impossible (due to a short nanorod length) or significantly slowed down at low temperatures.

It should be noted that even an infinite nanowire firstly breaks up into nanorods of different lengths, which may influentially determine the consequent scenarios of breakup. This is the reason to study nanowire dynamics for two types of boundary conditions: (i) with ′free′ ends (when the end-effect may develop immediately), and (ii) with 'frozen' ends, which is similar to periodic boundary conditions when, at least temporarily, the end-effect does not work till the first ruptures occur. In the second case, five atomic layers of motionless/frozen atoms are inserted between each end and the end-walls of the cylindrical container.

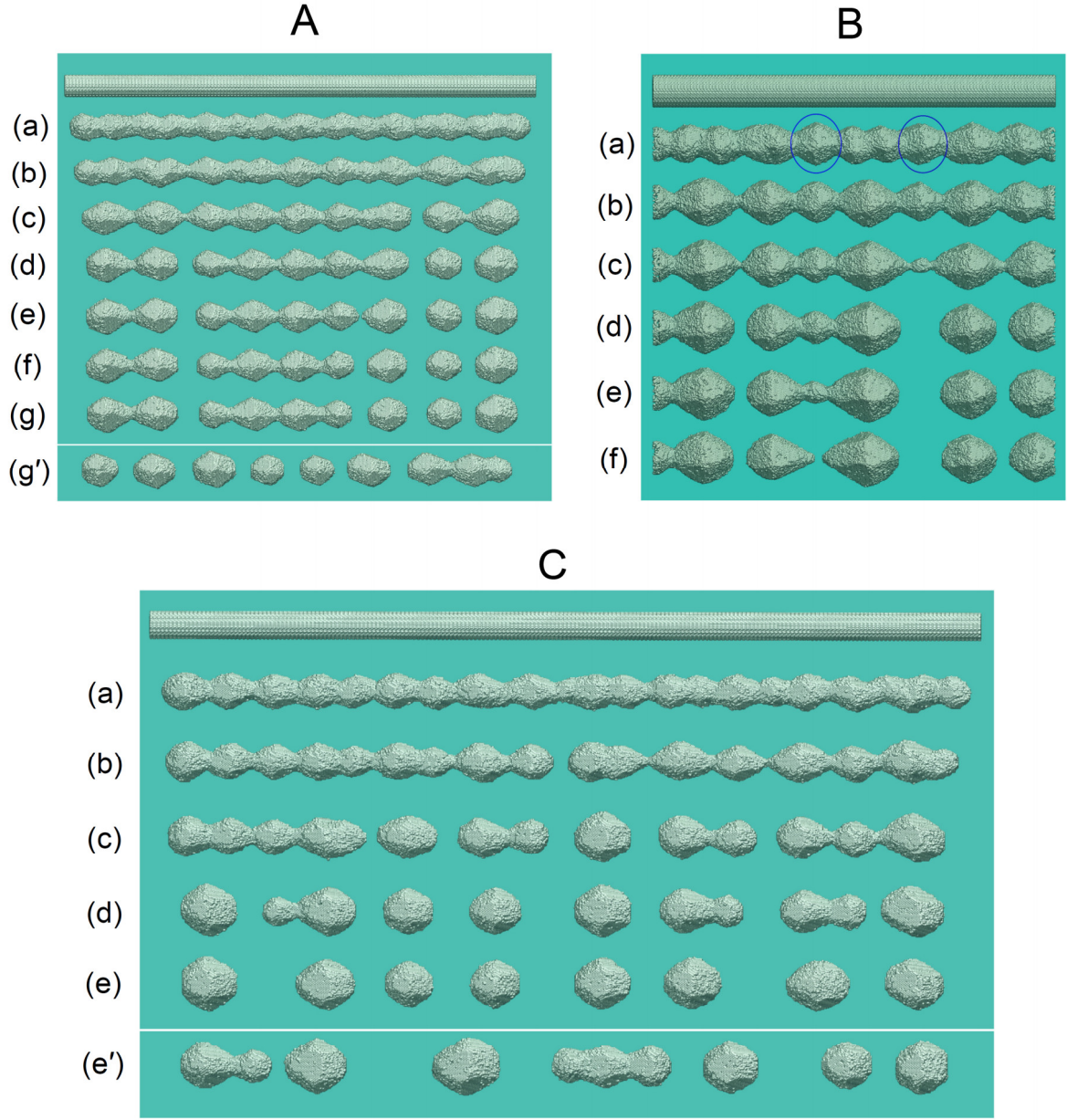

**Fig. 4.** (**A**) Dynamics of a nanowire at low temperature: $\alpha = 3.0, p = 0.32, L = 350$, and $d_0 = 16$. The initial nanowire is shown on the top of the figure; it contains $N_0 \approx 197 \times 10^3$ atoms, of which approximately $N_t \approx 192.2 \times 10^3$ atoms remained in the nanowire and nanoparticles after the formation of the gas. Sub-images (a), (b), (c), (d), (e), (f), and (g) show, respectively, the nanowire configurations after $t = 4, 8, 12, 16, 20, 24$ and $28\ (\times 10^6)$ MC time steps. Sub-image (g′) shows the system at the final stage of the breakup with a different random number sequence used starting from the initial configuration, $t = 25.1 \times 10^6$ MC time steps. (**B**) Breakup of a relatively thick nanowire at high temperature: $\alpha = 2.4, p = 0.40, L = 300$ and $d_0 \approx 25.4$, for which $N_0 \approx 398 \times 10^3$ and $N_t \approx 385.5 \times 10^3$. Sub-images (a), (b), (c), (d), and (e) correspond to the moments $t = 5, 15, 30, 36, 41$, and $42.6\ (\times 10^6)$ MC time steps, respectively. Blue ellipses represent the two beads that subsequently disappear with time. (**C**) Evolution of a nanowire at intermediate temperature: $\alpha = 2.7, p = 0.36, L = 500$, and $d_0 = 16.5$, for which $N_0 \approx 281.3 \times 10^3$ and $N_t \approx 268.8 \times 10^3$. Sub-images (a), (b), (c), (d), and (e) show, respectively, the nanowire configurations after $t = 3.5, 7, 10.5, 21$, and $28.4\ (\times 10^6)$ MC time steps. Sub-image (e′) depicts the system as it evolves with a different random number sequence used starting from the initial configuration, $t = 29.7 \times 10^6$.

In the nanowires with the diamond cubic crystal structure, the short-wave modulations of nanowire radius of large amplitude develop even under conditions when the end-effect manifestation is excluded (the ends of the nanowire are 'frozen' - Fig. 4B). Estimating the wavelength of the surface perturbations, $\lambda_{bead}$, in the configuration (b) of Fig. 4B, gives the value $\lambda_{bead}/r_{eff} \approx 3.5$, which is much lower than the threshold, $2\pi$, predicted by the classical model[39]. The effective radius of the nanowire,

$$r_{eff} = r_0\sqrt{N_t/N_0} \ , \tag{7}$$

characterizes its transverse size in quasi-equilibrium with the vapor of free atoms. The ratio $\lambda_{bead}/r_{eff}$ increases ($\lambda_{bead}/r_{eff} \approx 4.4$ for configuration (c) shown in Fig. 4A) with decreasing temperature.

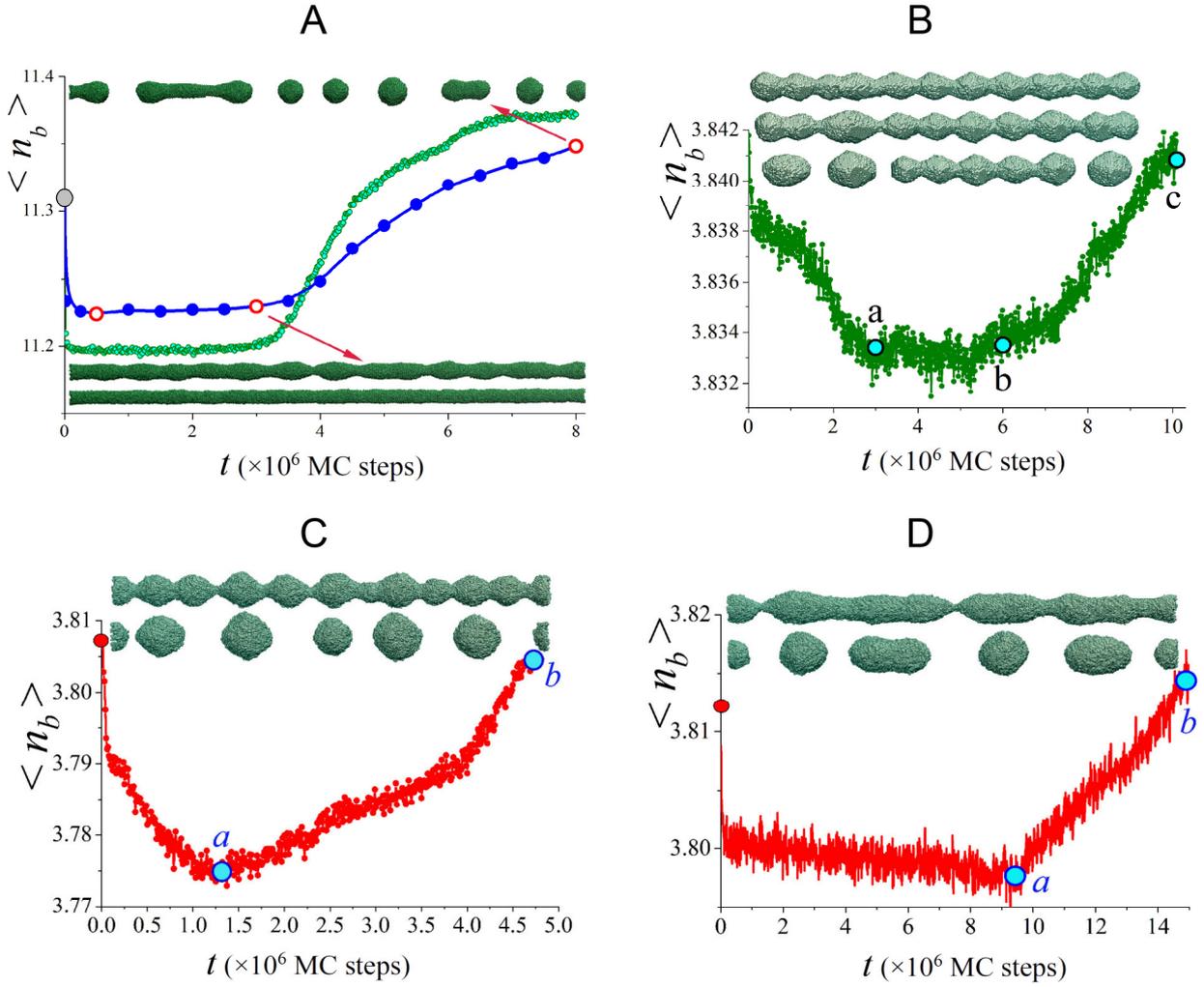

**Fig. 5.** Dynamics of the average number of bonds per one atom, $<n_b(t)>$, under different regimes of nanowire breakup. **(A)** The FCC crystal lattice, «warm» regime: $\alpha = 0.9$, $p = 0.725$, $L = 450a$, and $d_0 = 12a$ (here, $a$ is the FCC lattice constant). Green circles represent the $<n_b(t)>$ calculated using the model that takes into account exchanges by atoms between the nanowire surface and its own vapor. Blue circles/curve show the result in the case in which the evaporation is blocked, i.e. $P_{filter} = 0$. In the insets, the configurations of the nanowire at consecutive moments in time (from down to up) are marked by red circles. **(B)** Results for the diamond-like lattice at an intermediate temperature under the surface-vapor exchange: $\alpha = 2.7$, $p = 0.36$, $L = 300$, and $d_0 = 16$. The upper insets show the nanowire shapes (from up to down) at the moments marked by (a), (b), and (c) on the dependency $<n_b(t)>$. **(C&D)** Dynamics of the nanowire with frozen ends in hot regime ($L = 300$). In both cases, the numbers of bonded atoms in a quasi-equilibrium state with the surrounding vapor is around $N_t \approx 176 \times 10^3$, which corresponds to the effective diameter of the nanowires $d_{eff} \approx 16$. These results have been obtained using modified models when the random attempts to evaporate are partially blocked: $P_{filter} = 0.5$ **(C)**, and $P_{filter} = 0.1$ **(D)**.

The formation of numerous short-wave perturbations of the cross-section of a nanowire with substantial amplitudes cannot be completed with ruptures of the nanowire in all the regions of sausages. Simple estimates show that the total area (surface energy) of a set of spherical droplets will be smaller than the lateral surface area of a cylindrical nanowire (of the same volume) if each drop has been formed of the nanowire fragment with length $l_{cr} \geq 4.5 r_0$. This estimate is rather rough and does not take into account, for example, the anisotropy of the density of surface energy. However, it implies the existence of a lower limit for the size of the droplets formed.

Reduction in the number of droplets at the final breakup stage in comparison with the number of clearly formed beads (compare, for example, the number of beads in configuration (a) of Fig. 4C with configurations (e) and (e′) of the same figure) is realized either through the merger of neighboring beads into single large drops, or through the absorption of individual beads by the nearest neighbors. The first of these mechanisms is demonstrated by the results presented in Fig. 4C, obtained in the intermediate temperature regime ($\alpha = 2.7$). Furthermore, it is illustrated more expressively in Fig. S5 in the Supplementary Information, where the dynamics of a thicker nanowire is presented. The absorption of individual beads is predominantly inherent to high-temperature regimes, when the exchange by atoms between neighboring beads through the surrounding vapor is intensified, and the breakup parameter, $\lambda_{bead}/r_{eff}$, is minimal (Fig. 4B). At low temperatures, when the role of vapor in the transport of atoms decreases, the length $\lambda_{bead}$ increases and may exceed the critical breakup length $l_{cr}$ for a given nanowire radius. Then each element of the beads chain may transform into separate nanodroplet (configuration (g′) in Fig. 4A).

The demonstrated dependence of the breakup scenarios on temperature shows the important role of the vapor of free atoms at all the stages of nanowire dynamics. The formation of short-wave modulations of the nanowire cross-section is due to the well-known effect of roughening transition [41-46]. The sign of this phenomenon is that the morphology of the flat interface of a crystal that is in equilibrium with its vapor dramatically transforms at a particular temperature, $T_R$, followed by the appearance of steps, terraces, pits, and other structures. The magnitude $T_R$ depends on the surface orientation. Particularly, for Si (11m)-facets (m = 9, 7, and 4), the roughening transition temperature decreases monotonously with the increase of angles from the [001]-orientation[65]. However, the experimental results show that the (110) facet is kinetically roughened less than the (113), but the (111) facet is most resistant to the effect discussed[65,69].

Thus, the complicated physical situation arises if the surface of a nanowire, which is "isomerically rounded" at the initial stage of its evolution, is formed of a set of facets with different orientations. A rather rare specimen is the nanowire with [111]-orientation, which could be symmetrically bounded by six [110]-type faces that are parallel to the nanowire axis. Nanowires with such an orientation are the least resistant against the formation of short-scale beads that finally results in a set of nanodroplets (see Figs. 1, 4, and S5; and Movies S4 and S5). In outline, mechanisms responsible for the excitation of the nano-beads chain are discussed in the

Supplementary Information on the basis of simple qualitative physical models. The main features of this phenomenon are presented in Fig. 5.

The role of the roughening transition in the breakup of a nanowire depends not only on the orientation of the nanowire axis, but also on the type of its crystal structure. For example, the breakup of the FCC nanowires with [100]-orientation (Fig. 5A; the model parameters correspond to those used in previous studies[32,33]) is almost independent of the concentration of the surrounding vapor. The dynamics of the internal energy of the nanowire is reflected by the time dependence of the average number of bonds, $<n_b(t)>$, per one atom (in the FCC lattice the maximum number of nearest neighbors is 12).

In Fig. 5A one can see that the value of $<n_b(t)>$ after a short and insignificant decrease remains almost unchanged up to the formation of a well-outlined chain of beads to the time $t \approx 3 \times 10^6$ MC steps in both variants of nanowire dynamics (see green and blue curves in Fig. 5A when sublimation from the nanowire surfaces was blocked). In our numerical simulations, the concentration of free atoms was sometimes artificially changed in the following fashion. If the utilized MC model determined any atom to become free, then its detachment would be realized with probability $P_{filter} < 1$ (i.e., an additional probabilistic 'filter' was used). In the case under consideration, the variation of the parameter $P_{filter}$ does not lead to any noticeable change of value $\lambda_{bead}/r_{eff}$ (for the nanowire shown by the lower red arrow in Fig. 5A, $\lambda_{bead}/r_{eff} \approx 6.3$).

Recall that decreasing $<n_b(t)>$ corresponds to increasing the internal energy during the roughening transition. However, the free energy of the system at a constant temperature decreases due to the growth of its entropy. Initial decrease in $<n_b(t)>$ in Fig. 5A does not lead to a radical reconstruction of the surface. It is merely a result of the thermal roughening of the initially smooth nanowire surfaces when the characteristic height of surface perturbations is much smaller than the nanowire radius.

The results in Fig. 5B show a typical correlation between the dynamics of the $<n_b(t)>$ dependence and change in the morphology of the nanowire surface because of a pronounced effect of the roughening transition. The decrease in the average number of bonds takes a rather long time since a significant short-scale modulation of the cross section of the nanowire is associated with the displacement of a considerable mass. That the cause of these modulations is precisely the interaction of the nanowire surface with the surrounding vapor we confirm by the results presented in Fig. 5C and D obtained for different values of $P_{filter}$. Note that the variation of $P_{filter}$ in the range from 1 to 0.5 does not reduce the number of beads joining the most outlined chain. The results shown in Fig. 5C ($P_{filter} = 0.5$) present the typical nanowire configurations formed in the simulations for $1 \gtrsim P_{filter} \gtrsim 0.5$. (See also 'Movie S5', which presents entire dynamics of the nanowire). Conversion of the short-scale cross-section modulations into the long-wave surface perturbations is observed only with significantly reduced vapor concentrations (see Fig. 5D, $P_{filter} = 0.1$).

## 3.3. Regularities and irregularities in disintegrating nanowire morphology

When the nanowire is oriented along the [111]-axis, its breakup, as shown in the previous section, occurs at all values of the parameter $\alpha$ (temperatures) in the model used. Turning the nanowire axis up (along the [110]-axis) or down (along the [112]-orientation) leads to a fundamental change in the dynamics of the nanowire surface. Partially, this is due to the fact that in such orientations the [111]-type facets partly form the lateral surface of the nanowire at the initial stage of its evolution (Fig. 3b and d). Recall that according to the experimental data[65], such facets are known for their resistance to the roughening transition.

In the [110]-orientation, short-wave surface perturbations (like notches on the trunk of trees) occur only on the opposite edges formed by the pairs of [111]-facets (Fig. 3d). For the intermediate temperature, $\alpha = 2.7$, the breakup of such a nanowire (at least with a diameter of $d \gtrsim 6\,nm$ for Si) was not observed in our simulations. Even the end-effect does not pinch off nanodrops from the nanowire ends (Fig. S9). Only with an increase in the temperature, the effect of the roughening transition leads to the formation of short-scale, snake-like, disordered structures (Fig. S10; $\alpha = 2.4$), the numerous elements of which form a much smaller number of nanodrops during the stage of disintegration accompanied by the shrinking of cut-off fragments.

Nanowires with the [112]-orientation demonstrate unusual dynamics at intermediate temperatures, $\alpha = 2.7$. The facets of the lateral surface of (113)-type (Fig. 3b) are the most susceptible to roughening transition as compared with the (100)- and (110)- types [65,69]. The slopes of the notches can be formed by the fragments of more stable facets (see the upper inset in Fig. 6B; the blue and yellow contours mark the (100)- and (110)-fragments), which intersect the nanowire axis at relatively small angles of ~24° and ~16°, respectively. (Recall that the formation of new surfaces in the area of sausages leads to an effective reduction in the total surface area only at small inclination angles from secant planes relative to the axis of the nanowire).

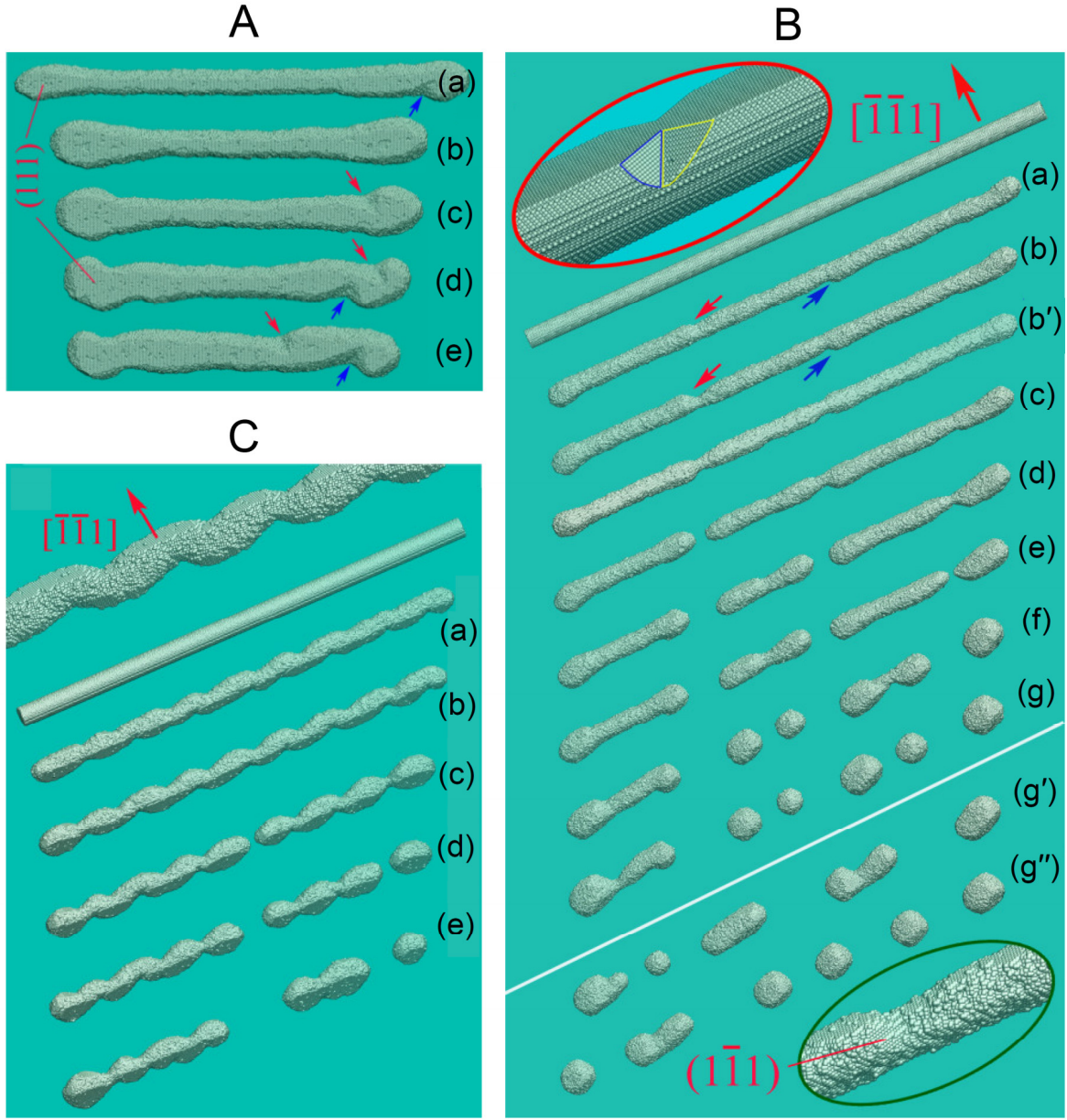

Fig. 6. (**A**) Dynamics of a nanowire oriented along the (112)-axis; warm regime: $\alpha = 2.7, p = 0.36, L = 450,$ and $d_0 = 16$. In configurations (a) to (d), $t = 41, 65, 70, 85,$ and $95 (\times 10^6)$ MC steps, respectively. (**B**) The [112]-orientation; hot regime: ($\alpha = 2.4$, $p = 0.4$, $L = 500$, and $d_0 = 17$. In configurations (a) to (g), $t = 4, 5, 6, 7.5, 8, 12$ and $15 (\times 10^6)$ MC steps, respectively. The initial nanowire is shown on the top, which contains approximately $N_0 \approx 318.6 \times 10^3$ atoms of which approximately $N_t \approx 278.5 \times 10^3$ atoms remained in the nanowire after the establishment of the gas. (b') denotes the view of configuration (b) from below. Arrows point to the original notches. The top/bottom insets illustrate the formation of notches on the surface of a nanowire at the initial stage. Configurations (g') and (g'') are the results of two other analogous numerical experiments; in both cases, $t = 13 \times 10^6$ MC steps. In configuration (g) the leftmost nanocluster is breaking up into two drops. (**C**) Break-up of a nanowire oriented along the bisector between vectors [112] and [1-10] at intermediate temperature: $\alpha = 2.7; p = 0.36$. $L = 400$ and diameter $d_0 = 16$. (a)-(e) - $t = 6, 10, 17, 22$ and $40 (\times 10^6)$ MC steps.

On the top and bottom of the nanowire, the directions of notches (i.e., directions from the (100)-fragments to the adjacent (110)-fragments, which form these notches) are opposite (for instance, see configuration (a) in Fig. 6B). At a relatively low temperature, notches develop only near the broadening ends of finite nanowires (see Fig. 6A; notches near the left end develop on the invisible, opposite face of (111)-type). A pair of symmetrical notches, formed simultaneously near one of the bounding (111)-faces, as shown in the upper inset in Fig. 6B, sometimes leads to a cut-off of the drop from the end of the nanowire. However, this symmetry is unstable. Instead, non-static bends of the nanowire ends without detachment from nanoclusters are more often observed. Single notches alternately appearing on the opposite sides of the nanowire (from top and bottom in Fig. 6A) drift toward its ends, which results in their bends up (configurations (c) and (d) of Fig. 6A) or down (configurations (a) and (e) of Fig. 6A). This phenomenon is a direct analogue of the well-known garden hose instability. In the case of nanowires, the driving instability factor is the excessive Laplace pressure on the tops of the rounded ends.

An increase in temperature (see Fig. 6B where $\alpha = 2.4$) leads to roughening transition on the entire lateral surface of the nanowire. Its initial stage is determined by asymmetrical notches discussed above (configurations (a) and (b) of Fig. 6B). If the intervals between adjacent neck regions are identified with the average period of the initial surface perturbations, $\Lambda$, then the results presented in Fig. 6B, Fig. S11, and Movie S6 give an estimate of $\Lambda/r_{eff} \sim 15$, which is very different both in length and in the shape of perturbations from the predictions of the model by Nichols and Mullins[39]. However, the number of droplets at the final stage of breakup largely exceeds the supposed value, $L/\Lambda$, because the end-effect mechanism breaks up the short fragments of the nanowire into separate drops (see configurations (d) and (g) in Fig. 6B).

It can be concluded that, for the [110]- and [112]-orientations, the nanowire breakup occurs only at high temperatures, $\alpha = 2.4$, on scenarios that are significantly different visually, and without explicit orderliness in the morphology of the nanowire surface. Below, we demonstrate one more example indicating that by changing the orientation of the nanowire, we could obtain nano-sized structures for various practical applications.

Suppose that, a nanowire axis is oriented along the direction that is the bisector between the [112]- and [1-10]-orientations. The nanowire, rounded at the initial stage of evolution, will still be bounded from 'top' and 'bottom' by facets the ($\mp 1 \mp 1 \pm 1$), i.e. by a pair of (111)-type facets (see the upper inset of Fig. 6C); however, its 'lateral' surfaces will be formed by the fragments of high-index facets and, accordingly, higher surface energy densities (with the susceptibility to roughening transition). The facets with lower surface energy densities (of the (110)-and (100)-types) intersect the nanowire axis at an angle of 24° (for the ($\pm 1 \pm 10$)-facets), and at a extremely small angle of $\approx 12°$ (for the ($0 \pm 10$)-facets). The pairs of these planes (i.e., (010) ∪ (110) and (0 − 10) ∪ (−1 − 10)) can form deep asymmetric notches on the opposite sides of the nanowire. The combination of these factors facilitates the development of a roughening transition on the surface of such a nanowire at lower temperatures ($\alpha = 2.7$) than in the cases of the [112]- and [110]-orientations ($\alpha = 2.4$). Moreover, at the stage of developed transformation, the nanowire surface represents an ordered, snake-like structure (see the top insert and configuration (b) in Fig. 6C and Movie S7 in the Supplementary Information). The relative period of this structure is $\Lambda/r_{eff} \approx 5.5$. The origin of this low value is the roughening transition.

In conclusion, we analyze the scenario of the breakup of the nanowire when its axis is oriented along the [001]-direction. In this case, the short-scale modulation of the nanowire cross-section cannot develop (Fig. 7), which originates from the following qualitative estimations. As it was mentioned earlier, the slopes of the supposed short-length beads have a tendency to be fragments of (111)-facets, while the lateral surfaces of these beads are presented by facets of (110)- and (100)- types (see Fig. 3a). In this assumed case, there would be explicit inhomogeneity of the surface energy density (the rate of sublimation) along the nanowire, with substantially lower values at the necking zones. Consequently, the dominating diffusion flows of free atoms from the necking regions to the vicinal thickening regions that are responsible for the breakup 'engine' do not arise. However, such flows may emerge for long-wave surface perturbations (see Supplementary Section S1.3) when the morphology of the surfaces of rather elongated nanowire fragments with different radii is similar to some extent. Then, both the surface diffusion and the spatial diffusion transport atoms from narrow to thick regions, which lead to the nanowire sausage instability as can be seen in Fig. 7. The parameter of the breakup, $\Lambda/r_{eff}$, essentially exceeds the classical value[39] (~9), with a proper estimation yielding the value of $\Lambda/r_{eff} \sim 18$ (see Movie S3 in the Supplementary Information).

At an intermediate temperature ($\alpha = 2.7$), no noticeable surface modulations were observed up to $t = 10^8$ MC steps. That is the reason that the parameters used in Fig. 7 ($\alpha = 2.4$) correspond to a Si-nanowire with a diameter of $\approx 4.2\ nm$ to decrease the time of simulations. The obtained results are in good agreement with experimental data[14]: if the nanowire temperature decreases from $T_1 \approx 1175^0$ K to $T_2 \approx 1050^0$ K, then the time of breakup, $t_{br}$, increases from 3 minutes to 14 hours. By the way, the ratio $T_1/T_2 \approx 1.12$ is almost the same as the ratio of used parameters, $\alpha$ : $2.7/2.4 = 1.125$.

The huge increase in $t_{br}$ usually occurs even in the hot regime if the evaporation of atoms from the nanowire surface is blocked ($P_{filter} = 0$). Thus, the exchange by atoms between the nanowire surface and surrounding gas of free atoms results in significant deviation of the parameter of breakup, $\Lambda/r_{eff}$, from the well-known value of ~9, either to short-length beads at the initial stage (Figs. 4, 6C, and S10) or to long-wave perturbations (Figs. 6B and 7).

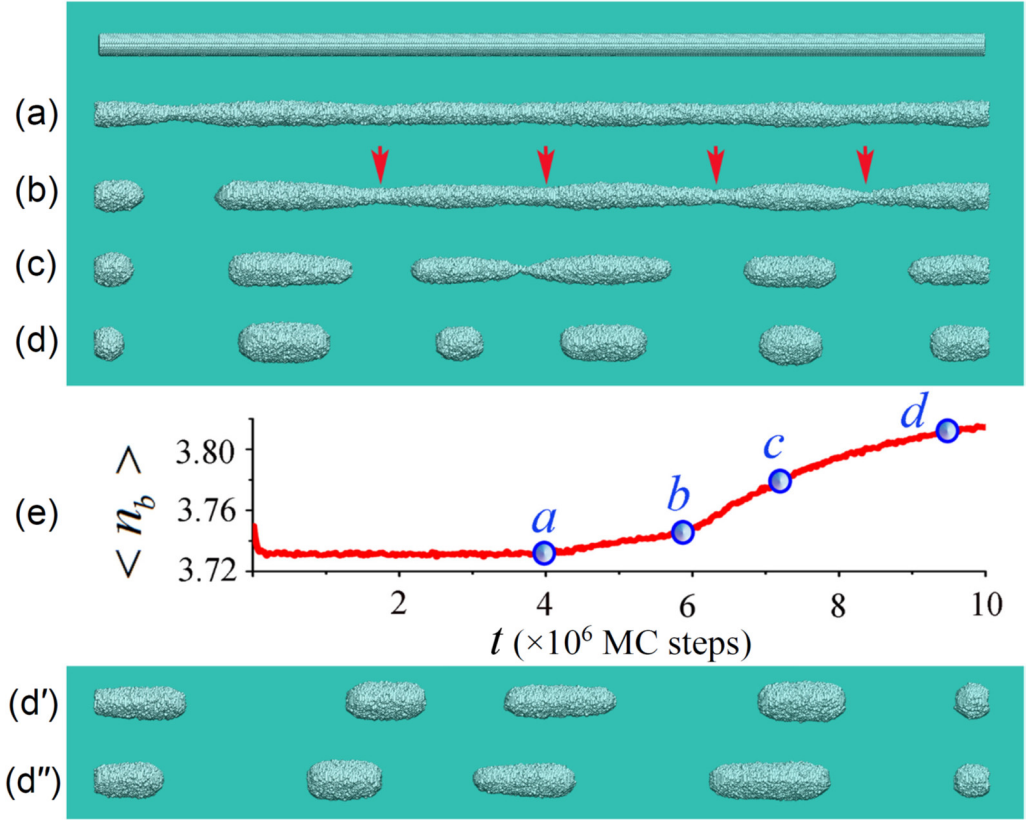

**Fig. 7.** Dynamics of a [100]-nanowire at high temperature: $\alpha = 2.4$, $p = 0.40$, $L = 500$ and $d_0 = 12$. The initial nanowire is shown on the top, which contains $N_0 \approx 155.7 \times 10^3$ atoms, of which approximately $N_t \approx 131 \times 10^3$ atoms remained in the nanowire and nanoparticles after the establishment of the gas ($r_{eff} \approx 5.5$). Sub-images (a), (b), (c) and (d), show, respectively, the nanowire configurations after $t = 4, 5.8, 7.2, 9.5$ ($\times 10^6$) MC time steps. Sub-images (d′) and (d″) show the nanowire shape at the final stage of breakup with a different random number sequence used. Subpart (e) shows the dynamics of the average number of bonds per one atom, $< n_b(t) >$. Points (a), (b), (c) and (d) on the graph indicate $< n_b(t) >$ for four nanowire configurations (a), (b), (c), and (d) depicted above, respectively.

## 4. Discussion and Conclusions

The features of the breakup of nanowires with a diamond crystal lattice structure - analyzed in the previous section - are not analogous with the breakup of nanowires with an FCC lattice structure. Furthermore, these features cannot be described on the basis of the classical model[39]. The difference in the number of nearest neighbors, $m_c$, and in the topology of the structure of interatomic bonds, determine the degree of anisotropy of the surface energy density for nanoclusters of the same shape. Moreover, an additional factor appears that significantly affects the dynamics of the nanowire, which is the exchange by atoms between the surface of the nanowire and the surrounding vapor of free atoms. The effect of restructuring the surface at a temperature $T > T_R$ (i.e., the effect of roughening transition) has been experimentally studied for Silicon in two-dimensional systems[44,45], where the extended crystalline facets of different orientations were the objects of research.

A nanowire is often considered to be a one-dimensional physical system, and such an approximation is sufficiently accurate in some cases. However, based on the one-dimensional approximation, it is impossible to describe the dynamics of a nanowire when its lateral surface is formed by longitudinal strips/facets with different orientations and with different susceptibilities to roughening transition. If the axis of the nanowire does not coincide with the axis of symmetry of its crystal structure, then such inhomogeneities lead to short-scale bends of the nanowire and fractures on its surface (Fig. 6, and Figs. S9 and S10). Note that the time-varying bends (up and down) of the end of the nanowire with [112]-orientation (Fig. 6A) during its contraction can be due to the same physical factors that lead to periodic deviations in the direction of growth of the synthesized nanowire away from the general Si [112]-direction[70].

The breakup dynamics of nanowires oriented along the [111]- and [100]- axes of the third- and fourth-order of symmetry, respectively, are characterized by an enormous difference in wavelengths, $\hat{\lambda}$, of the developed modulations of the nanowire radius from $\hat{\lambda}/r_{eff} \sim 4$ to $\hat{\lambda}/r_{eff} \sim 18$. Both of these values are inconsistent with the traditional notion of Plateau-Rayleigh instability[39]. The first one is noticeably lower than the critical value, $2\pi$, which is necessary for the development of instability, and the second one is as twice as the well-known ratio, $\hat{\lambda} \approx 9r_0$. The reason for such discrepancies between the results obtained and the classical model is that this model did not take into account the role of the vapor formed by the nanowire itself at high temperatures. It is the transport of free atoms from one region of the nanowire to the other that substantially affects the dynamics of the nanowire breakup. The corresponding physical mechanisms - in a rather crude approximation - are quantitatively analyzed in section S1.1 of the Supplementary Information; it turned out that the associated results are close to those obtained using our Monte Carlo model.

In outline, a qualitative description of the physical factor that determines the breakup of a nanowire is as follows. Assume that the surface energy density, $\sigma$, is isotropic. The nanowire radius varies along the longitudinal coordinate, $z$, as $\Delta r = -\varepsilon \cos(2\pi z/\hat{\lambda})$. The modulation amplitude, $\varepsilon$, can increase in time if at the narrowing zone ($\Delta r \lesssim 0$) the flux density of evaporating atoms, $\Gamma_{vap}(z)$, dominates the flux density of atoms returning to the surface, $\Gamma_{dep}(z)$, from the common vapor, ($\Gamma_{vap}(z) > \Gamma_{dep}(z)$), and at the broadening region ($\Delta r \gtrsim 0$) the inverse relation holds ($\Gamma_{vap}(z) < \Gamma_{dep}(z)$).

It is easy to establish that the above conditions are satisfied in the case of long-wave perturbations, $\hat{\lambda} \gtrsim 15 r_0$, for which the surface curvature, $\kappa$, depends chiefly on the radius of the nanowire cross section, $\kappa \approx r^{-1}(z)$. In some approximation, the quantity $\Gamma_{vap}(z)$ is proportional to the local curvature of the surface, whereas the distribution $\Gamma_{dep}(z=0)$ is an integral function of this curvature, and to a lesser extent, depends on the local value of the radius of the nanowire. Thus, the monotonic functions of $z$, $\Gamma_{vap}(z)$ and $\Gamma_{dep}(z)$, satisfy inequalities

$$\Gamma_{dep}(z=0) > \Gamma_{dep}(z = \hat{\lambda}/2), \quad \Gamma_{vap}(z=0) > \Gamma_{vap}(z = \hat{\lambda}/2), \tag{8}$$

and

$$\Gamma_{dep}(z=0)/\Gamma_{dep}(z = \hat{\lambda}/2) < \Gamma_{vap}(z=0)/\Gamma_{vap}(z = \hat{\lambda}/2). \tag{9}$$

(It should be noted that the first inequality in Eq.(8) is an unfavorable factor for the development of instability.)

Since the nanowire is in a quasi-equilibrium state with the surrounding vapor, in which the number of free atoms remains practically unchanged, then, given Eqs. (8) and (9), this state can be realized only under conditions

$$\Gamma_{dep}(z \sim 0) - \Gamma_{vap}(z \sim 0) < 0 \quad \text{and} \quad \Gamma_{dep}(z \sim \lambda/2) - \Gamma_{vap}(z \sim \lambda/2) > 0, \tag{10}$$

which correspond to the development of instability. Note that in the case under consideration, surface diffusion fluxes are also directed from the regions of narrowing of the nanowire to the regions of broadening.

For short-wave perturbations, $\hat{\lambda} < 2\pi r_0$, in Eqs. (8) and (9), the signs of the inequalities are reversed, leaving the pair of relations in Eq. (10) valid. However, the surface diffusion factor in such cases inhibits the development of instability, limiting the spectrum of possible values of $\hat{\lambda}$ from below.

The anisotropy of the surface energy density determines the length of the developing perturbations of the nanowire depending on both its orientation and the type of crystal lattice, fitting this length with the basic law of thermodynamics: the free energy of the nanowire at a given temperature should not increase during its transformations. In the case of the FCC lattice, estimates of the parameter $\hat{\lambda}/r_{eff}$ made on the basis of experimental data (by the number of droplets formed) differ from the classical value, ~9, by no more than 30%. Such deviations make it possible to assert that the experimental results are in good agreement with the theoretical model. Indeed, the role of vapor and anisotropy of σ may slightly affect the number of droplets and the breakup time in some cases (as can be seen from the FCC lattice in Fig. 5A). However, nanowires with a diamond lattice structure clearly demonstrate the features of the manifestation of their crystal structure: depending on orientation, either 'ultra-shortwave' or 'extra-longwave' modulation of the nanowire cross section can be excited. In the first case, the further dynamics of the beads (formed at the stage of nanowire breakup) and the formation of single nanodroplets is a direct analogue of the well-known Ostwald ripening mechanism (neighboring beads absorb each other; see Fig. 4B and Movie S5 in the Supplementary Information). In the second case, the number of periodic modulations of the surface is usually equal to the number of droplets formed (see Fig. 7 and Movie S3 in the Supplementary Information).

The results of our numerical experiments are in good agreement with previously obtained experimental data[12,14] (Fig. 1 and Fig. S6). However, for the complete confirmation of the main conclusions of our theoretical findings, additional experimental studies are required.

## Acknowledgments

V. Gorshkov and V. Tereshchuk acknowledge the support of the Ministry of Education and Science of Ukraine (Project N 2904-f). V. Gorshkov would like to thank his late colleague Prof. V. Privman of Clarkson University for the productive discussions they had in the course of analysis of the data and preparation of the manuscript.

## Appendix A. Supplementary data

Supplementary data (including text and videos) to this article have been submitted along with the main manuscript.

## Data availability

The raw/processed data required to reproduce these findings cannot be shared at this time due to technical or time limitations.

# Supplementary Information

**Restructuring and Breakup of Nanowires with the Diamond Cubic Crystal Structure into Nanoparticles**

Vyacheslav N. Gorshkov[1], Vladimir V. Tereshchuk[1], and Pooya Sareh[2*]

[1]National Technical University of Ukraine, Igor Sikorsky Kyiv Polytechnic Institute, 37 Prospect Peremogy, Kiev 03056, Ukraine.

[2]School of Engineering, University of Liverpool, London Campus, 33 Finsbury Square, London EC2A 1AG, United Kingdom.

*Corresponding author. Email: pooya.sareh@liverpool.ac.uk

## Appendix A: Qualitative analysis of the mechanisms of roughening transition of the nanowire surface

### S1. Excitation of the ultra-shortwave and extra-longwave modulations of the nanowire cross-section

**S1.1.** In this section, we present the effect of vapor of free atoms around a nanowire on the wavelength of the developing perturbations on the basis of a simplified physical model. Suppose that free atoms in the diffusion mode deposit on the surface of the nanowire, the radius of which varies periodically along its axis, $z$, as follows

$$r(z) = \sqrt{1 - \varepsilon^2/2} - \varepsilon\cos(kz), \quad k = 2\pi/\lambda. \quad (S1)$$

Here, $\varepsilon$ is the amplitude of the radius modulations, and $\lambda$ is the wavelength of the perturbations (in Eq. S1), and in the sequel, the initial radius $r_0(z, \varepsilon = 0) = 1)$. The dependence of the radius on the coordinate $z$, Eq. S1, corresponds to the preservation of the volume of the wire with the change of amplitude, $\varepsilon$. The boundary conditions for the concentration, $n$, of free atoms are as follows: at the infinity, $n(\infty) = n_0$, and on the surface of the nanowire, $n|_{surf} = 0$.

In this approximation, Fig. S1 shows the distribution of the relative density of the diffusion flux, $\eta(z) = \Phi(z)/\Phi(z = \lambda/2)$, on the surface of the nanowire along a section of length $\lambda/2$.

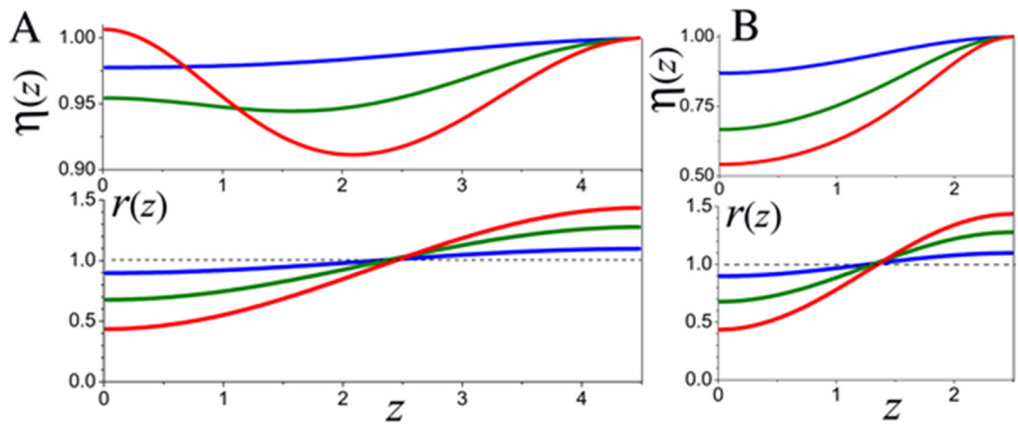

**Fig. S1.** Distributions of the relative density of the diffusion flux, $\eta(z) = \Phi(z)/\Phi(z = \lambda/2)$, and the surface profiles of the nanowire for $\varepsilon = 0.1, 0.3, 0.5$ (blue, green and red curves, respectively). **(A)** $\lambda = 9$. **(B)** $\lambda = 5$.

One can see that for the wavelength $\lambda = 9$ (Fig. S1A), the diffusion flux density in the narrowing region exceeds its value in the broadening region ($\eta(z = 0) > 1$) if the perturbations of the radius become sufficiently large. For $\lambda = 5 < 2\pi$ ($\lambda = 2\pi$ is the lower limit of unstable perturbation modes according to the models [38,39]), the diffusion flux density in the narrowing region decreases monotonically as the amplitude of perturbations increases (Fig. S1B). Such a dependence $\eta(z = 0, \varepsilon)$ can lead to the excitation of short-scale instabilities, with wavelengths values below the classical limit $\lambda = 2\pi$ or $k = 1$.

For a more accurate determination of this range, it is necessary to take into account the imbalance between the flows of atoms to the surface and the flows of evaporating atoms. For a nanowire with isotropic physical surface characteristics, the evaporation flux density in some approaches is proportional to the Laplace pressure or its relative value $\hat{p}_L(z) = p_L(z)/p_L(z = \lambda/2)$. For a cylindrical surface, we have

$$p_L \sim \frac{1}{r(z)\xi} - \xi^{-3}\frac{d^2 r(z)}{dz^2}, \quad \xi = \left[1 + \left(\frac{dr(z)}{dz}\right)^2\right]^{1/2}. \tag{S2}$$

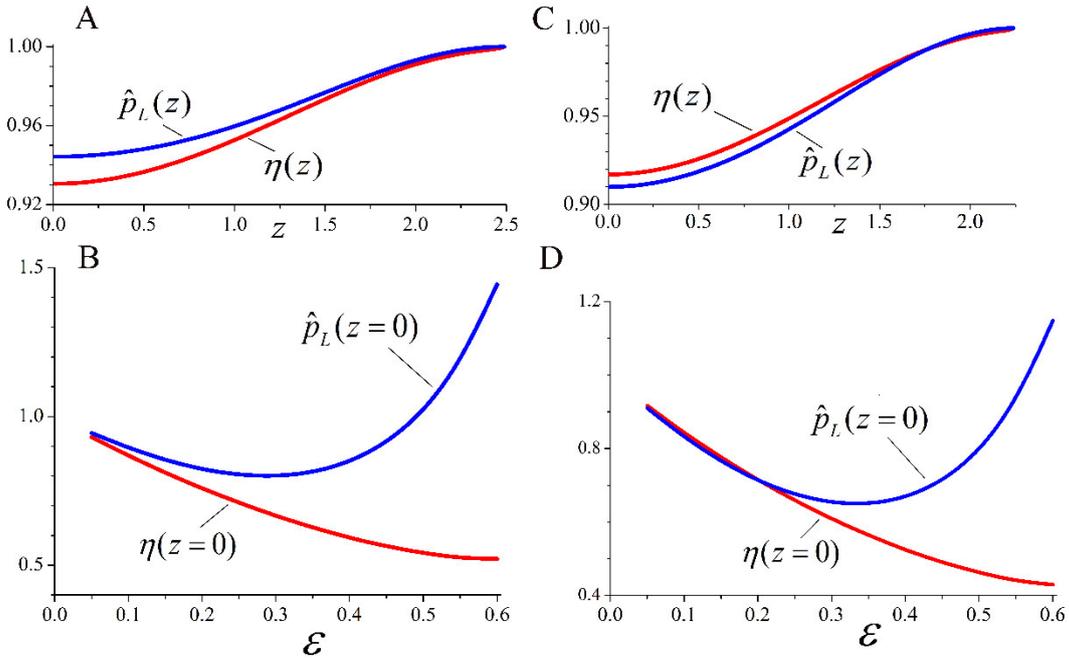

**Fig. S2.** (A & C) Distribution of relative flux densities, $\eta(z)$, and relative surface pressures, $\hat{p}_L(z)$, along the surface of the nanowire with $\varepsilon = 0.05$ for (A) $\lambda = 5$, and (C) $\lambda = 4.5$. (B & D) Dependencies of $\eta(z)$ and $\hat{p}_L(z)$ in the zone of constriction ($z = 0$) on the amplitude of the perturbation, $\varepsilon$, of the nanowire radius.

In the self-consistent equilibrium of the nanowire surface with its own vapor, the number of free atoms is approximately constant during the break-up process. In other words, the flux of atoms on the surface of the wire and the sublimation flux from the surface are equal to each other

$$\int_0^{\frac{\lambda}{2}}[\eta(z) - \gamma\hat{p}_L(z)]r(z)\,\xi(z)dz = 0, \tag{S3}$$

where $\gamma$ is the coefficient of proportionality between the surface pressure $\hat{p}_L(z)$ and the sublimation flux density, the value of which must be found from the integral Eq. S3. The initial perturbations of the nanowire radius will increase if, for the found parameter $\gamma$, the flux density, $\varphi(z) = \eta(z) - \gamma \hat{p}_L(z)$, is negative in the narrowing region and positive in the broadening region. This condition is guaranteed to be satisfied if the curve $\hat{p}_L(z)$ lies above $\eta(z)$.

According to the data from Fig. S2A, the instability with the wavelength $\lambda = 5$ can indeed develop since the gap between the monotonic dependences $\hat{p}_L(z)$ and $\eta(z)$ expands as the perturbation $\varepsilon$ increases (Fig. S2B). For a shorter wavelength, $\lambda = 4.5$, (Fig. S2C), perturbations with a small amplitude should be attenuated. Such a mode can increase in time only for initial perturbations $\varepsilon \gtrsim 0.25$, where $\hat{p}_L(z = 0) > \eta(z = 0)$ (Fig. S2D). The distribution of the flux density, $\varphi(z)$, at $\lambda = 5$ and $\varepsilon = 0.05$ is shown in Fig. S3 (corresponding value $\alpha = 0.9936$).

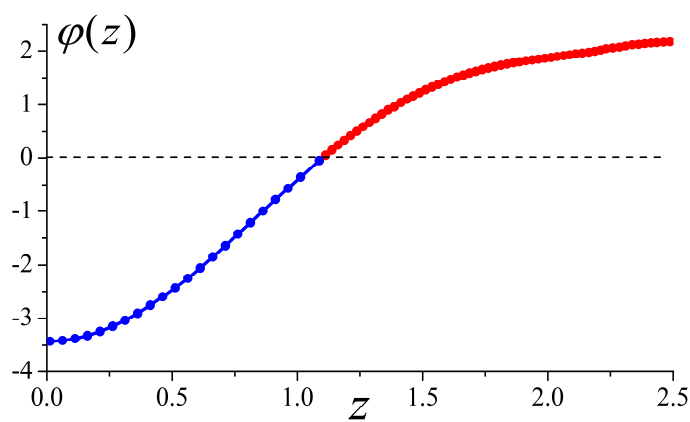

**Fig. S3.** Distribution of the flux density, $\varphi(z)$, on the surface of the nanowire at $\lambda = 5$ and $\varepsilon = 0.05$. The values of $\varphi(z)$ are given in arbitrary units.

Thus, we have shown only the possibility of the development of short-wave perturbations of a nanowire with a wavelength $4.5 < \lambda < 2\pi$ based on a simple mathematical model and qualitative estimates without taking into account the factor of surface diffusion of atoms. At $\lambda < 2\pi$, the surface diffusion from the broadening zones (with a lower binding energy of surface atoms) to the narrowing zones may inhibit the development of instability.

The role of the exchange of atoms between the nanowire surface and its own vapor (Fig. S3) depends on the average concentration of free atoms and can dominate only with the temperature increasing. In conclusion, we note that the anisotropy of the surface energy density and the fact that surface perturbations are not exactly sinusoidal (Eq. S1) make significant corrections to the dynamics of the nanowire disintegration, as was shown by the results of our numerical experiments.

**S1.2.** The results presented in Fig S4 illustrate the role of the factors discussed above. The parameter $\alpha$ is chosen to be small, which corresponds to the high temperature and high rate of the surface and spatial diffusion of atoms. The rate of evaporation of atoms from the surface is regulated in our model as follows. If, according to the initial rules of the model, an atom has a chance to break away from the surface of the nanowire, then an additional ′filter′ confirms the realization of this detachment with probability $P_{filter}$. Such a technique allows one to clearly demonstrate the peculiarities of the role of vapor in the break-up process.

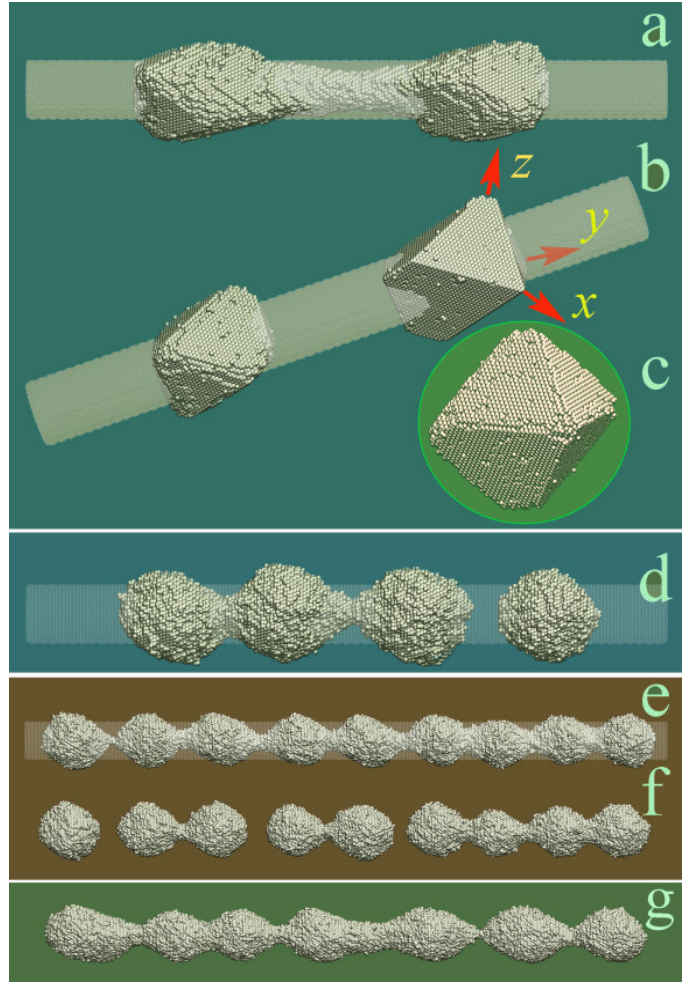

**Fig. S4.** Effects of exchange of the nanowire surface with the surrounding vapor of free atoms. High temperature: $\alpha = 1.8$ and $p = 0.5$. Sub-images (a) and (b) show the transformation of the nanowire in the case in which evaporation is blocked, $L = 160$, $r = 7$, the number of atoms is 68400, $t = 55$ and 81 ($\times 10^6$ MC steps), respectively. Sub-image (c) depict the formation of the right cluster at the moment $t = 10^8$. In sub-image (d), $L = 160$, $r = 9.8$, evaporation is taken into account, $r_{eff} \approx 7.2$ (a cylinder with such a radius is delineated in the figure), $t = 10^6$, and the number of atoms is 73100. In sub-images (e) and (f), $L = 290$, $r = 9$, the evaporation probability is reduced by 50%, $r_{eff} \approx 8.15$ (a cylinder with this radius is shown in the figure), and $t = 1.2$ and 2 ($\times 10^6$ MC steps), respectively; the number of atoms is 170300. In sub-image (g), the evaporation probability is reduced by 80%, $r_{eff} \approx 8.3$, $t = 2.8 \times 10^6$ MC steps, and the number of bound atoms is 178800.

If the evaporation of atoms from the surface of the nanowire is blocked ($P_{filter} = 0$), then periodic surface perturbations are not observed. Break-up occurs only due to the so-called end-effect[33] (sub-images (a), (b), and (c) of Fig. S4). The formed nanodroplets are in the form of octahedrons (see the main text). When $P_{filter} = 1$ (sub-image (d) of Fig. S4), short-wave surface perturbations arise. Reducing the value of $P_{filter}$ to 0.5 does not change the linear density of the 'beads' (sub-images (e) and (f) of Fig. S4). However, a significant increase in the wavelengths of perturbations is noticeable if the rate of evaporation decreases, $P_{filter} = 0.2$ (sub-image (g) of Fig. S4). Thus, the density of the beads depends non-linearly on the parameter $P_{filter}$, which indicates the threshold nature of the roughening transition.

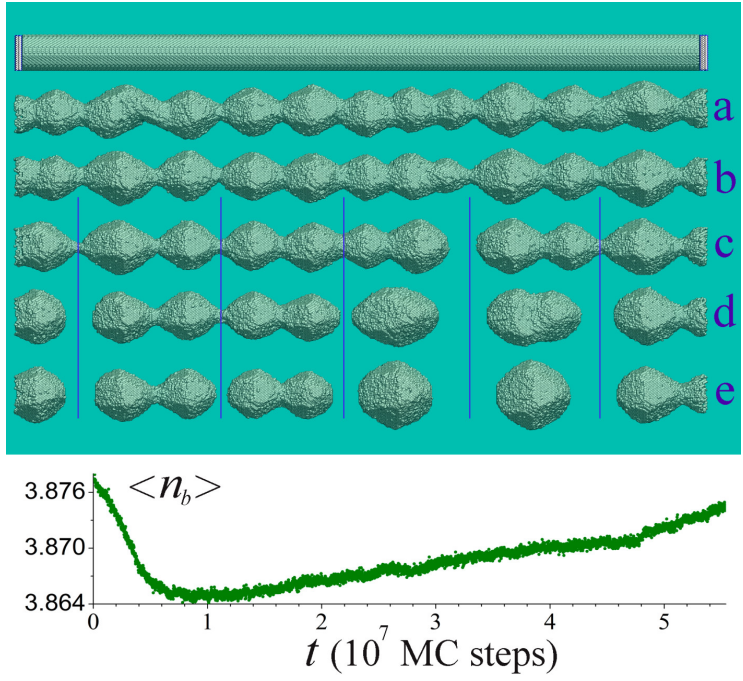

**Fig. S5.** Fragmentation of the nanowire in the process of breakup according to the scenario of merging pairs of adjacent beads: $\alpha = 2.7, p = 0.36$, $d = 21, L = 400$, $N_0 \approx 388 \times 10^3$ (after the establishment of equilibrium with gas $N_t \approx 376.7 \times 10^3$ and $d_{eff} = 20.7$), and $t = 10, 16, 24, 48, 55.3$ ($\times 10^6$ MC steps), respectively. On the top of the figure, the initial shape of the nanowire with layers of 'frozen atoms' is shown. The graph on the bottom shows the dynamics of the average number of bonds per atom, $< n_b(t) >$.

Therefore, a specific feature of the diamond-like lattice is the possibility of initiating anomalously short-wave perturbations on the nanowire surface ($\lambda \sim (4 - 4.2) \times r_{eff}$) at the initial stage of its dynamics. However, it is impossible to preserve such a short-scale structure down to ruptures in all the original neck regions according to the laws of thermodynamics. The formation of enlarged nanoclusters can occur in different ways.

One option is to combine pairs of neighboring beads into single drops, which is demonstrated in Fig. S5. Here we consider the dynamics of a thicker nanowire than the one shown in Fig. 4C (see the main text) at the intermediate temperature. An increase/decrease in temperature changes the length of the developed modulations of the cross-section size of the nanowire and, thereafter, the mechanism for the formation of gaps (see the explanation of Fig. 4 in the main text).

**S1.3.** In the previous section, we discussed the role of the gas of free atoms when chains of the short-length beads form. If in our qualitative estimations the wavelength, $\lambda$, increases, the dependencies $\eta(\lambda; z, \varepsilon)$ and $\hat{p}_L(\lambda; z, \varepsilon)$ become non-monotomic functions of $z$ (as can be seen from Fig. S1A, for instance). This effect initiates the excitation of irregularities in the nanowire dynamics and may lead to decreasing the instability increment at the given $\lambda$. However, these irregularities can be suppressed when $\lambda$ exceeds the classical value 9 ($r_0(z, \varepsilon = 0) = 1$; see Eq. S1). Such a possibility can be realized, for instance, at $\lambda = 16$ (see Fig. S6A). However, this qualitative estimation seems to be rather justified; the parameter of nanowire breakup, $\Lambda/r_{eff} \sim 18$, was obtained when simulating the nanowire dynamics with the [100]-orientation (Fig. 7 and Fig. S6B). It is seen that our results are in good agreement with the experimental results in ratios among the average values of transverse sizes of the nanowire fragments, their lengths, and the distances between the centers of vicinal nanoclusters. It should be noted that, in this case, the surface diffusion and the spatial diffusion transport atoms in the same direction – from the neck regions to the thickening regions. Thus, the vapor of free atoms may either excite the short-scale beads, or stimulate longwave perturbations depending on the nanowire orientation and according to the laws of thermodynamics.

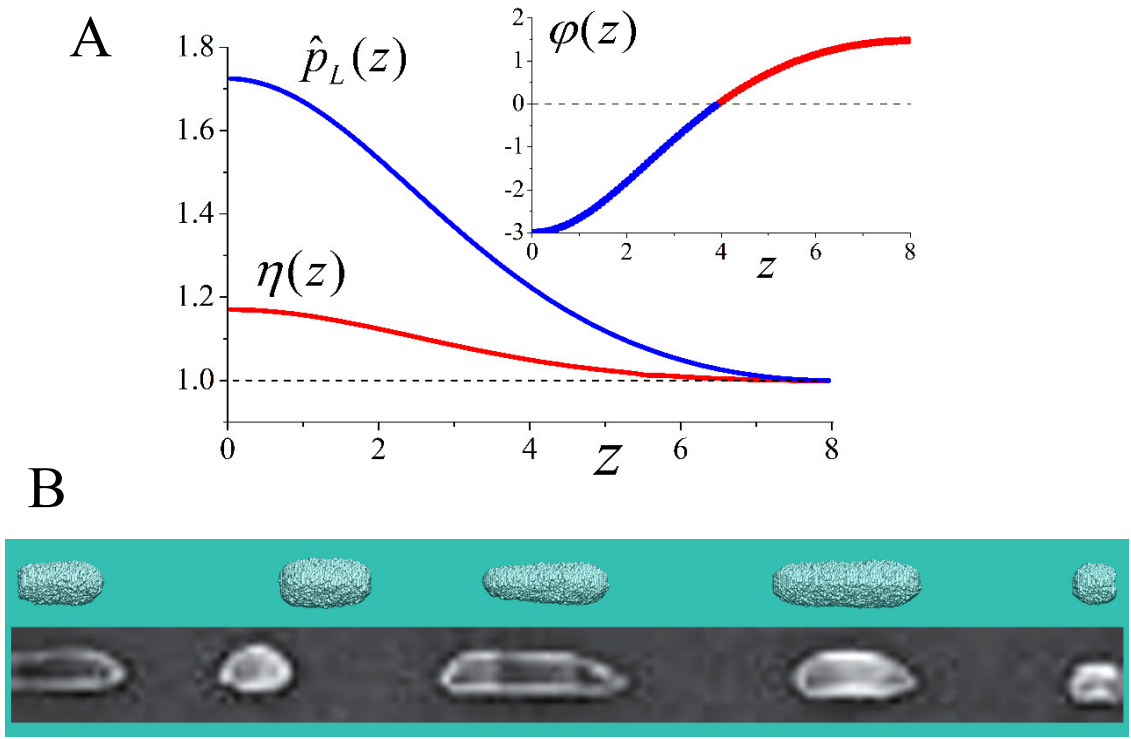

**Fig. S6.** (**A**) Distributions of relative flux density, $\eta(z)$, and relative surface pressure, $\hat{p}_L(z)$, along the surface of the nanowire with $\varepsilon = 0.3$ and $\lambda = 16$. In the inset, the distribution of the corresponding flux density, $\varphi(z)$, is shown (in arbitrary units) which satisfies Eq. S3 at $\gamma \approx 0.852$. (**B**) Fragments of nanowires obtained from a real experiment[14] and from our numerical simulations for the [100]-nanowire.

A consequence of the large value of the breakup parameter, $\Lambda/r_{eff} \sim 18$, is that finite nanorods with the [100]-orientation may be rather stable (Fig. S7). For example, in the initial configuration shown in Fig. S7A, the length of the nanorod, $L$, is around $60r_0$; however, the visible signs of the breakup are not observed while the breakup occurs in the [111]-, [110]- and [112]-orientations, even if the ratio $L/r_0 \lesssim 40$ (Figs. S5, S10, and S11).

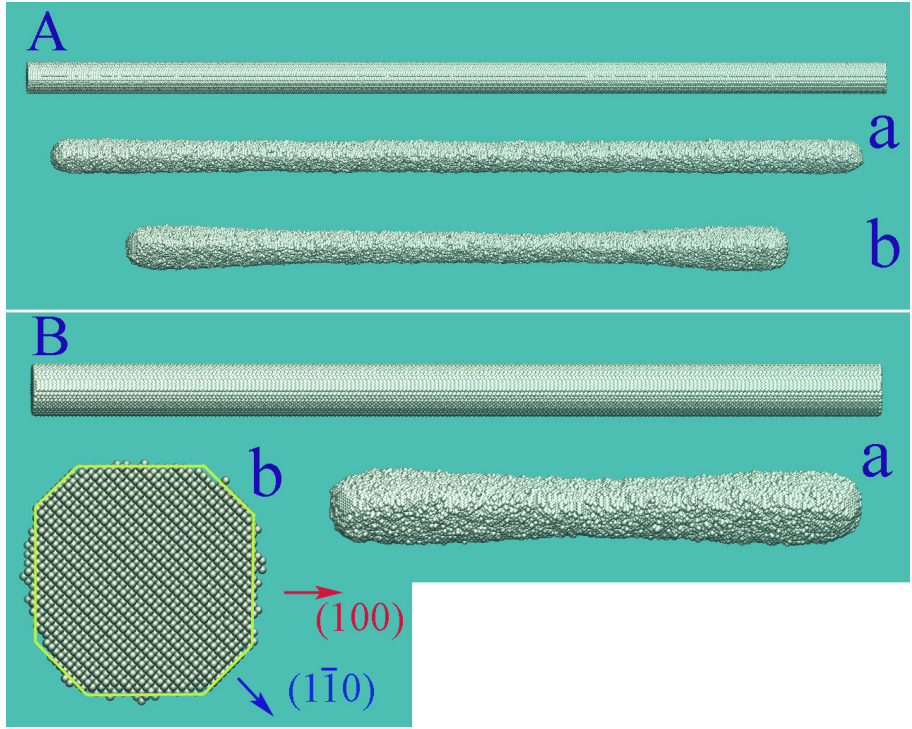

**Fig. S7.** Dynamics of the finite nanorods with the [100]-orientation. **(A)** $d = 17, L = 500$. In (a), the nanowire shape is achieved by the time $t = 17 \times 10^6$ at low temperature ($\alpha = 3$ and $p = 0.32$); In (b), the result is shown at high temperature ($\alpha = 2.4$ and $p = 0.4$). **(B)** Dynamics of a shorter nanowire at high temperature: $d = 17, L = 300$, and $t = 18.5 \times 10^6$. In sub-image (b), a cross-sectional view of the nanowire in the central part is depicted. As a result of isomeric rounding, its lateral surface is bounded by (100)- and (110)-type planes.

## S2. Generation of long-lived fragments

It is important to point out that during the process of nanowire break-up, relatively short fragments are frequently formed with the number of beads of 2-3-4[31-33], which remain in the quasi-equilibrium state for a long time and resemble the Delaunay unduloids[66]. The subsequent evolution leads either to disintegration into separate nanodroplets, or to merger into a large cluster. For example, Fig. S8 shows the formation of a 3-body dumbbell (depicted within the blue rectangle), which forms at $t = 12 \times 10^6$ MC time steps and stays in the quasi-equilibrium state for a long time. The ambiguity of its final form manifests itself at $t \gtrsim 42 \times 10^6$ MC time steps. Both breakup into two separate parts (configuration (f′) in Fig. S8), and very slow merging into a single whole, from the state shown in configuration (f), can occur. The formation of a similar 3-body quasi-equilibrium configuration from a short Si-nanorod with a diameter of $\approx 9.2 \, nm$ nm is shown in Movie S2.

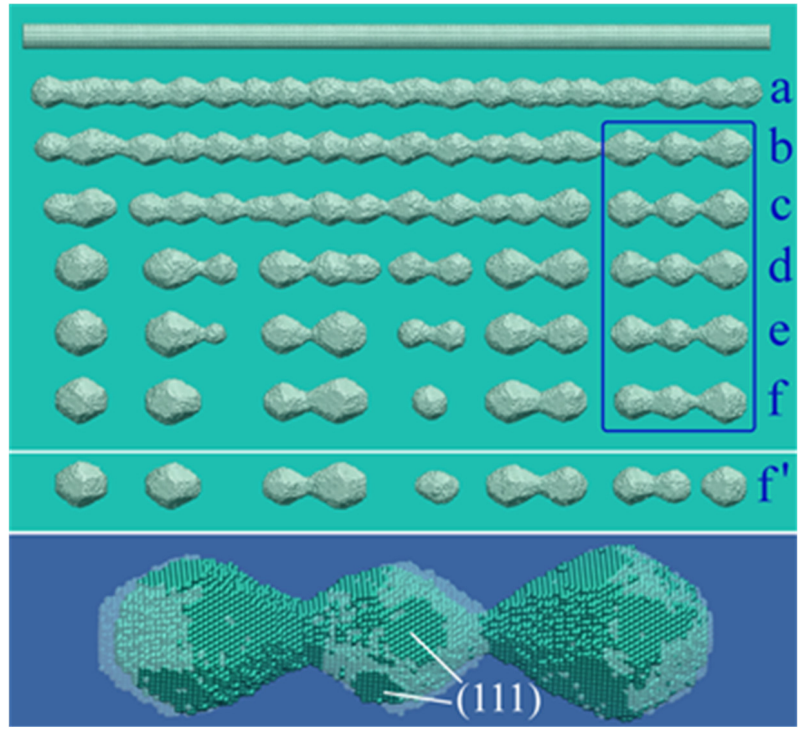

**Fig. S8.** Nanowire of an initial [111]-orientation, with length $L = 500$ and diameter $d = 16$, evolved according to the MC dynamics at low temperature value corresponding to $\alpha = 3.0$ and $p = 0.32$. The initial nanowire, shown on the top, contained approximately 281.3 atoms, of which approximately 275.0 atoms remained in the nanowire and nanoparticles after the establishment of the gas. Sub-images (a), (b), (c), (d), (e) and (f) show, respectively, the nanowire configurations after $5, 11, 17, 26, 36$, and $50$ ($\times 10^6$ MC time steps). Sub-image (f′) shows the system as it evolves with a different random number sequence used starting from configuration (c), $t = 47 \times 10^6$ (the separation of the rightmost drop occurred at $t \approx 45 \times 10^6$). The apparently stable three-headed dumbbell on the right formed at time $12 \times 10^6$ MC steps. It is magnified in the bottom inset, in which the transparent olive balls denote atoms in it at time $17 \times 10^6$ MC steps, whereas the solid balls show the atoms at time $37 \times 10^6$ MC steps.

### S3. Non-axisymmetric perturbations of the nanowire surface

The neck slopes of a nanowire with the [111]-orientation are formed by planes of [100]- and [111]-type orientation. Moreover, the axis of the nanowire - with a good approximation - is the axis of symmetry of the 3rd order (see, for instance, Fig. 4B, Fig. S8 (the lower inset), and Fig. S5). However, even in this orientation, the fragments of a nanowire with helicoidal and snake-like morphology often appear (see Movie S1 and Movie S4). Figs. S9-S11 demonstrate (i) the threshold nature of the development of the roughening of a nanowire surface with the temperature increasing, and (ii) the axial symmetry breaking of the nanowire at the nonlinear stage of break-up and the formation of zigzag-like structures.

In the intermediate thermal regime ($\alpha = 2.7$), the break-up of the nanowire with the [110]-orientation was not detected even for extremely large simulation times (Fig. S9). The concentration of the vapor of free atoms is not sufficient for an intense exchange of atoms between the notches and pyramidal protrusions on the lateral surface of the wire (notches on the nanowire from above and below are formed by [100]-type faces (Fig. 3d). The temperature increase ($1/T \sim \alpha = 2.4$) 'activates' the roughening transition (Fig. S10).

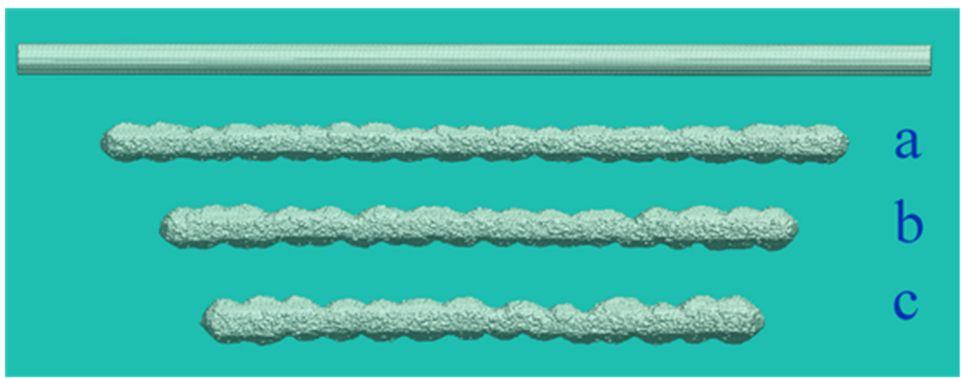

**Fig. S9.** Nanowire of an initial [110]-orientation, with length $L = 500$ and diameter $d = 16$, evolved according to the MC dynamics at intermediate temperature corresponding to $\alpha = 2.7$ and $p = 0.36$. The initial nanowire, shown on the top, contained approximately 280.3 atoms, of which approximately 263.0 atoms remained in the nanowire after the establishment of the gas. Sub-images (a), (b), and (c) show, respectively, the nanowire configurations after $40, 80,$ and $120$ ($\times$ $10^6$ MC time steps).

The minimization of the free energy during the deformation of the nanowire corresponds to the fact that the notches at the top and bottom are shifted relative to each other; the nanowire bends appear, and the irregularly shaped nanoclusters are formed (configuration (e) in Fig. S10A and configuration (f) in Fig. S10B). At the initial stage of the break-up, the wavelength, $\hat{\lambda}$, of periodic perturbations is of the order $(4 - 4.2) \times r_{eff}$.

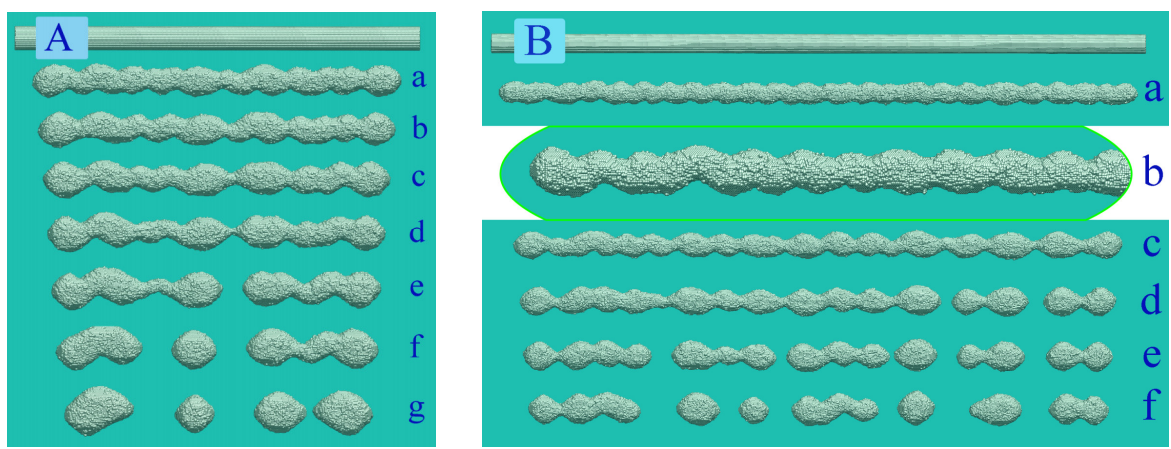

**Fig. S10**. A series of time-snapshots of MC runs for a nanowire along the [110]-type lattice direction ($\alpha = 2.4$ and $p = 0.40$). (A) $d = 17$ and $L = 300$; from the top to bottom, the MC times are $t = 4, 6, 8, 10, 12, 14, 18$ ($\times$ $10^6$ MC steps). The initial nanowire is shown in the first image. (B) $d = 16.5$ and $L = 600$; $t = 2, 4, 7, 10, 12, 15$ ($\times$ $10^6$ MC steps). The shape (b) presents only the left part of the nanowire.

In the case of [112]-orientation, asymmetric notches from above and below arise to a large extent chaotically. This randomness can be seen not only in Fig. 6A but also in the additional results presented in Fig. S11 and Movie S6. An estimate of the characteristic distance, $\hat{\lambda}$, between the adjacent neck regions at the initial stage of break-up gives, in this case, the value $\hat{\lambda} \approx 15 r_{eff}$. If we assume that the characteristic length of perturbations, $\hat{\lambda}$, is equal to the ratio of the length of the nanowire to the average number of droplets formed ($<n_{drop}> \approx 6$), then we obtain $\hat{\lambda} \approx 10.5 r_{eff}$. The noticeable difference in these results exists because the formed, extended fragments of the nanowire continue to break up due to the end-effect. Hence, the estimate of $\hat{\lambda}$ by the number of droplets per unit length is applicable only to the spontaneous disintegration of an inviscid liquid jet. In the case of a nanowire with a crystalline structure, such an evaluation usually leads to a distorted view of the physical mechanisms of its breakup.

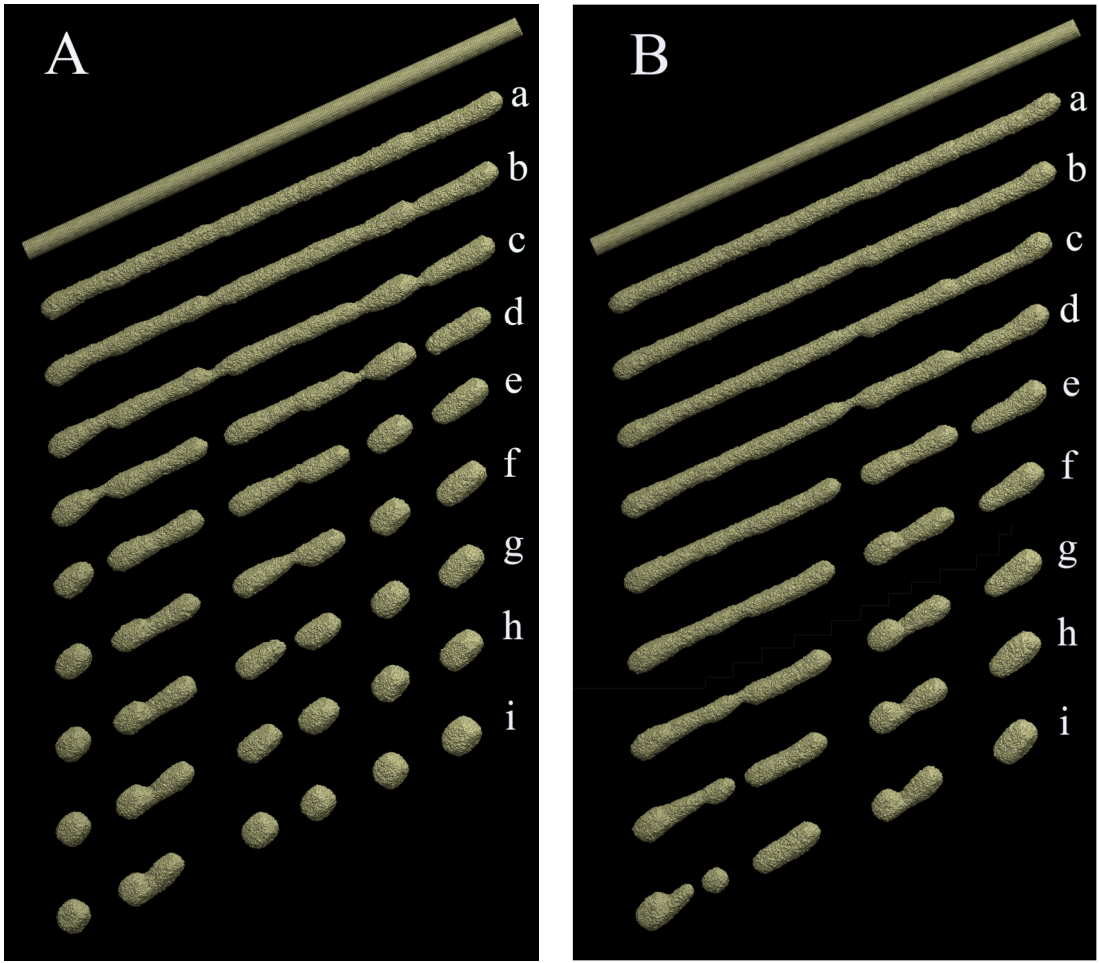

**Fig. S11.** Two random nanowire dynamics with the [112]-orientation. Hot regime ($\alpha = 2.4$, $p = 0.4$, $L = 500$, and $d_0 = 17$). The initial nanowire, shown on the top, contained approximately $N_0 \approx 318.6 \times 10^3$ atoms; after the establishment of the gas $N_t \approx 278.5 \times 10^3$. (**A**) In configurations (a) to (i), respectively, $t = 3, 4, 5, 6, 7, 8, 9, 10, 14$ ($\times 10^6$ MC steps). (**B**) In configurations (a) to (i), respectively, $t = 3, 4, 5, 6, 7, 8, 9, 10, 12$ ($\times 10^6$ MC steps ).

Movie S1

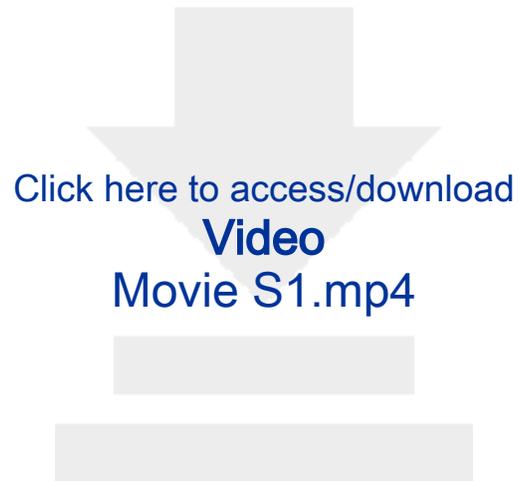

Click here to access/download
Video
Movie S1.mp4

Movie S2

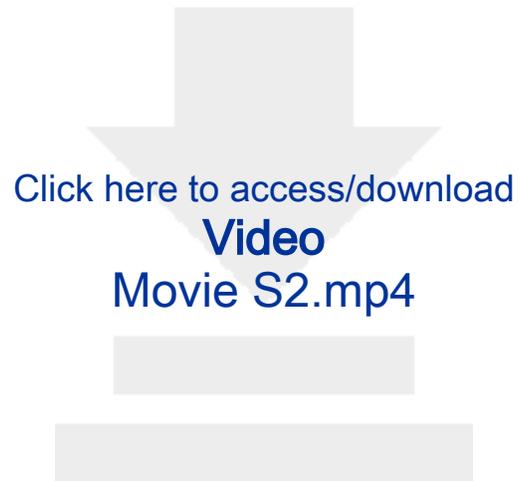

Click here to access/download
**Video**
Movie S2.mp4

Movie S3

Click here to access/download
Video
Movie S3.mp4

Movie S4

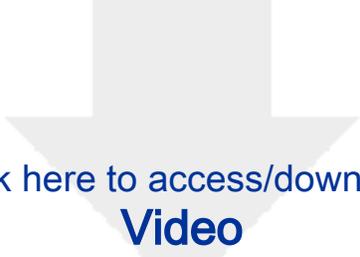

Click here to access/download
Video
Movie S4.mp4

Movie S5

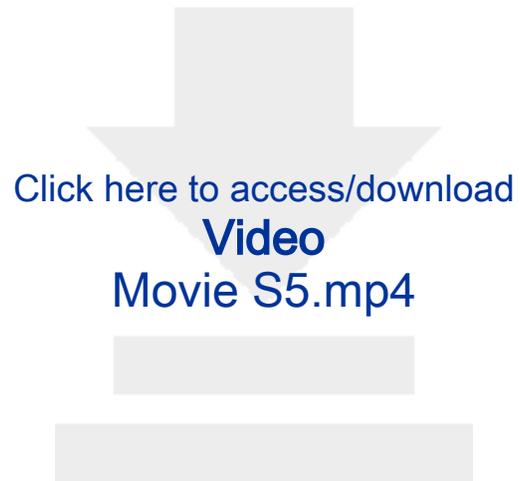

Click here to access/download
**Video**
Movie S5.mp4

Movie S6

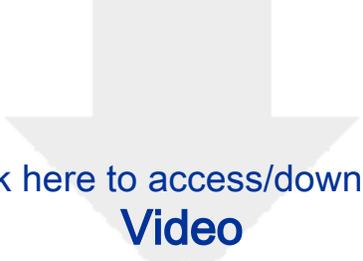

Click here to access/download
**Video**
Movie S6.mp4

Movie S7

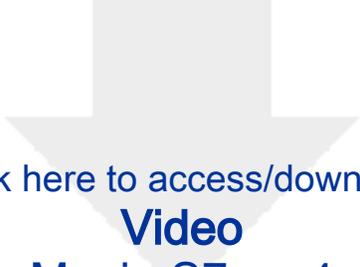

Click here to access/download
Video
Movie S7.mp4